%% file: AMultAniPlate.tex
\documentclass[1p]{elsarticle}
\biboptions{longnamesfirst,square,sort,compress}

\usepackage[T1]{fontenc}
\usepackage[latin1]{inputenc}
\usepackage[english]{babel}
\usepackage{graphicx}
\usepackage{enumerate}

\usepackage{amssymb,amsmath}
\usepackage{empheq}
\usepackage{mathtools}
\usepackage{mathrsfs} 
\usepackage{frcursive}
\usepackage{color}
\usepackage{comment}
\usepackage{multirow}
\usepackage{booktabs}
\usepackage{soul}
\usepackage[section]{placeins}

\newcommand{\mymat}[1]{
\mathbf{#1}
}
\everymath{\displaystyle}
\allowdisplaybreaks[4]
\usepackage{hyperref}
\hypersetup{
    unicode=true,          
    pdftoolbar=true,        
    pdfmenubar=true,        
    pdffitwindow=false,     
    pdfstartview={FitH},    
    pdfnewwindow=true,      
    colorlinks=true,       
    linkcolor=red,          
    citecolor=green,        
    filecolor=magenta,      
    urlcolor=cyan           
}
\begin{document}

\begin{frontmatter}
	\title{A multilayer anisotropic plate model with warping functions for the study of vibrations reformulated from Woodcock's work}
	\author[drive]{A.~Loredo\corref{cor1}}
	\ead{alexandre.loredo@u-bourgogne.fr}
	\author[drive]{A.~Castel}
	\ead{alexis.castel@u-bourgogne.fr}
	\cortext[cor1]{Corresponding author}
	%
	\address[drive]{DRIVE, Université de Bourgogne, 49 rue Mlle Bourgeois, 58027 Nevers, France}
	%

\begin{abstract}
In this paper, a suitable model for static and dynamic analysis of inhomogeneous anisotropic multilayered plates is described. This model takes into account the variations of the transverse shear strains through the thickness of the plate by means of warping functions. Warping functions are determined by enforcing kinematic and static assumptions at the interfaces. This model leads to: a $10\times10$ stiffness matrix coupling to each other the membrane strains, the bending and torsion curvatures, and the $x$ and $y$--derivatives of the transverse shear strains; and a classical $2\times2$ transverse shear stiffness matrix. This model has been proven to be very efficient, especially when high ratios between the stiffnesses of layers --up to $10^6$-- are present.
\par 
This work is related to Woodcock's model, so it can be seen as a reformulation of his work. However, it brings several enhancements: the displacement field is made explicit; it is reformulated with commonly used plate notations; laminate equations of motion are fully detailed; the place of this model among other plate models is now easy to see and is discussed; the link between this formulation and the original one is completely written with all necessary proofs; misses and errors have been found in the energy coefficients of the original work and have been corrected; it is now easy to improve or to adapt the model for specific applications with the choice of refined or specific warping functions. 
\par
Static deflection and natural frequencies for isotropic and anisotropic sandwich plates are given and compared to other models: they show that the present model is very accurate for the simulation of such structures.
\end{abstract}
\end{frontmatter}
\input{01_Introduction.tex}
\input{02_Model.tex}
\input{03_PlaceOfModel.tex}
\input{04_Conclusion.tex}
\section*{Acknowledgement} 
This work was supported by the \emph{Burgundy Region} and the \emph{European Regional Development Fund}.
\bibliographystyle{unsrt}
\bibliography{AMultAniPlate}
\appendix
\input{10_Appendix1.tex}
\input{11_Appendix2.tex}
\input{12_Appendix3.tex}
\label{dernierepage}
\end{document}

%% file: 01_Introduction.tex
\section{Introduction}
Static and dynamic behavior of plates have been studied by many authors for a long time. Among them, we can cite the early works of Kirchhoff~\cite{Kirchhoff1850}, Rayleigh~\cite{Rayleigh1944}, Love~\cite{Love1888}, achieved at the end of the 19th century, which are appropriate for the study of thin plates. They were followed by the works of Reissner~\cite{Reissner1945}, Uflyand~\cite{Uflyand1948}, Mindlin~\cite{Mindlin1951} in the middle of the 20th century, suitable for moderately thick homogeneous plates. At the same time, the behavior of anisotropic multilayered plates began to be studied by Lekhnitskii~\cite{Lekhnitskii1935} and Ambartsumyan~\cite{Ambartsumyan1958}.
\par
Since these works, static and dynamic behavior of anisotropic multilayered plates have been extensively studied, as it can be seen in Carrera's review article~\cite{Carrera2002}. In composite applications, the high ratio between shear and longitudinal stiffnesses and the strong inhomogeneities between layers make classical models inefficient. It has motivated researchers to develop more refined theories, able to handle the behavior of such structures.
\par
As done by the previously cited early works, these refined theories attempt to describe the displacement field in the thickness direction by means of kinematic and/or static hypothesis. The difference is that, in the recent works, this description is more refined. In particular, the transverse shear strain variations through the thickness have received many attention, as it can be seen in references~\cite{Pai1995,Kim2006,Brischetto2009,Vidal2011,Neves2012,Dozio2012}, among others. The more recent and pertinent theories have been built using \emph{warping functions} which take into account these variations. In these models, the first spatial derivatives of the transverse shear strains are coupled with the membrane and bending terms, leading to a more complex stiffness matrix of size $10\times10$ compared to the classical $6\times6$ stiffness matrix of the previous models.
\par
The model presented in this paper belongs to this last category as it will be shown later. It has been first formulated in an early work of Sun and Whitney~\cite{Sun1973}, it is the second model described in this reference. Even if the model is presented for the most general case, authors have established equations of the model only for symmetric isotropic laminates. This model was improved by other researchers~\cite{Guyader1978,Woodcock2004} to study the dynamic behavior of more general multilayered plates and also to simulate the sound transmission through these structures. This model has been proven to be very efficient for the study of the viscoelastic behavior of strongly inhomogeneous structures like plates damped with viscoelastic PCLD patches~\cite{Loredo2011,Castel2012}. All these works were limited to on-axis orthotropic plies. Woodcock has extended his first work to laminates including off-axis orthotropic plies~\cite{Woodcock2008}.
\par
The present work reformulates Woodcock's last model~\cite{Woodcock2008}. This reformulation has been motivated by the following disadvantages of the original work: 
\begin{itemize}[--] \topsep 0pt \itemsep 4pt \parskip 0pt
  \item the displacement field is not made explicit. It is only known implicitly and involves constraints on both displacement and stress fields;
	\item the notations used by the author are not the standard notations for plate models, which makes the work difficult to understand;
	\item some helpful formulas are not given and missing terms and errors have been found on the given energy coefficients;
	\item due to the complexity of the model and its formulas, implementation is difficult;
	\item links between this model and other plate models are not easy to foresee.
\end{itemize}
The reformulation of this work leads to the following enhancements:
\begin{itemize}[--] \topsep 0pt \itemsep 4pt \parskip 0pt
  \item the displacement field is now fully explicit. It involves four warping functions $\varphi_{\alpha\beta}(z)$;
	\item the global plate behavior is given in terms of classical (membrane / bending / shearing) generalized displacements, forces, and stiffnesses, using the most common notations of plate theories;
	\item the link between the present model and Woodcock's model is fully detailed;
	\item helpful formulas for computing Woodcock's coefficients $\alpha^{\ell}_{x},\alpha^{\ell}_{xy},\beta^{\ell}_{x}\dots$ are given;
	\item the corrected list of all the 69 Woodcock's strain energy coefficients $\lambda_i$ is given. It is shown that the list reduces to 55 independent coefficients in the most general case;
	\item a corrected and extended list (18 instead of 13) of Woodcock's coefficients $\delta_i$ for the kinetic energy is given; It is shown that the list reduces to 14 independent coefficients in the most general case;
	\item the place of this model among other plate theories is now easy to see, and is discussed;
	\item as the model only involves integrals through the thickness, it is easy to implement;
	\item the present reformulation makes improvements easy to implement: warping functions may be changed or enhanced without complicating the model.		
\end{itemize}
In addition, the static deflection and fundamental frequency of isotropic sandwich plates with various Young's modulus ratio between the layers are computed with the help of Navier's procedure. This study compares the present model's results to several other classical and more advanced models. A second numerical study gives natural frequencies for two anisotropic sandwiches issued from the literature. For both cases, results are also compared to a three dimensional finite element simulation.

%% file: 02_Model.tex
\section{Model}
\subsection{Laminate definition}
\begin{figure}
	\centering
		\includegraphics{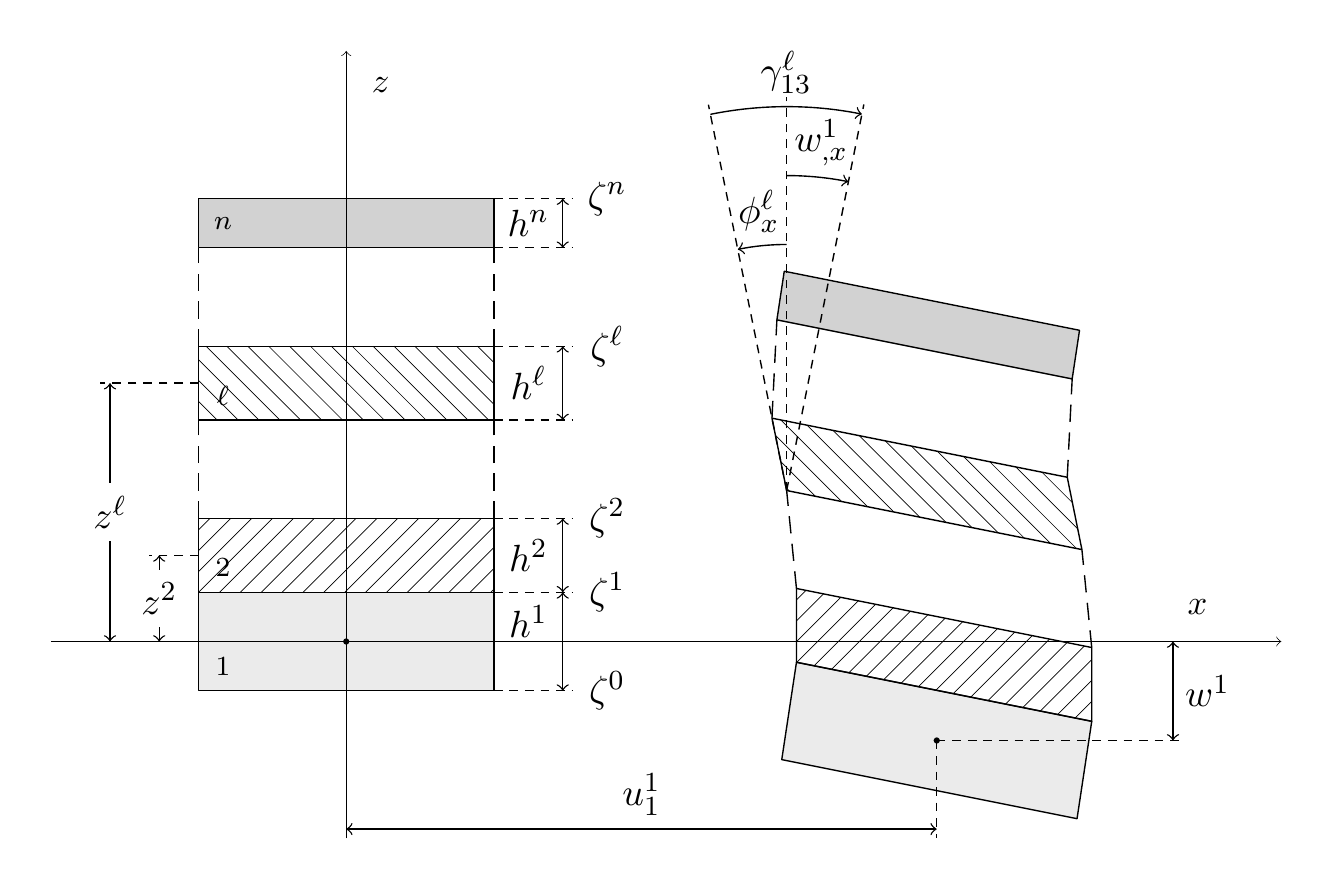}
	\caption{Geometrical parameters of the undeformed laminate on the left side, and its deformed shape with corresponding quantities on the right side.}
	\label{fig:Figure1}
\end{figure}

The laminate is composed of $n$ layers. In this study, all the quantities will be related to those of the first layer, which then plays a central role. Figure~\ref{fig:Figure1} illustrates the following definitions:
\begin{itemize}[--] \topsep 0pt \itemsep 4pt \parskip 0pt
	\item $z^{\ell}$ is the elevation/offsetting of the middle plane of the layer $\ell$
	\item the $\ell^\text{th}$ layer is located between elevations $\zeta^{\ell-1}$ and $\zeta^{\ell}$
\end{itemize}
With these definitions:
\begin{itemize}[--] \topsep 0pt \itemsep 4pt \parskip 0pt
	\item there are $n$ parameters $z^{\ell}$ with, by definition, $z^{1}=0$,
	\item there are $n+1$ parameters $\zeta^{i}$ with $i$ taking values from $0$ to $n$,
	\item the thickness of the layer $\ell$ is $h^\ell=\zeta^{\ell}-\zeta^{\ell-1}$
\end{itemize}
\par
Plate models classically take the laminate middle plane as reference plane, then the laminate is located between~$-h/2$ and $h/2$ where $h$ is the total height of the laminate. This model has been developed in a slightly different way: the middle plane of the first layer is the reference plane. This is particularly useful, although not essential, to study plates damped with patches~\cite{Loredo2011,Castel2012}. In this last case, the number of layers can vary from a point to another and the definition of a middle plane is difficult to establish. Of course, one can change the reference plane in this formulation without changing the behavior of the model, but all formulas have to be changed in consequence. 
\subsection{Kinematic and static assumptions of the model}
In the following, Greek subscripts take values $1$ or $2$ and Latin subscripts take values $1$, $2$ or $3$. The Einstein's summation convention is used for subscripts only. The comma used as a subscript index means the partial derivative with respect to the directions corresponding to the following indices.
\subsubsection{Displacement field}\label{sec:DisplacementField}
First, kinematic assumptions of Woodcock's model are reported below after a change of notations in order to agree with most of plate models notations. They are:
\begin{align}
\label{eq:depl_W}
  \left\{
    \begin{array}{lll}
      u_\alpha^\ell(x,y,z) & = & u_\alpha^\ell(x,y,z^\ell) + (z^\ell-z)(w^1_{,\alpha}(x,y)-\gamma^\ell_{\alpha3}(x,y)) \\ 
      u_3^\ell(x,y,z) & = & w^1(x,y) 
    \end{array} 
  \right.
\end{align}
where $\ell\in[1..n]$ is the number of the layer, $z^{\ell}$, $u_\alpha^\ell(x,y,z)$, $\gamma^\ell_{\alpha3}(x,y)$ and $u_3^\ell(x,y,z)$ are respectively the $z$-coordinate of the mid-plane, the in-plane displacements, the engineering transverse shear strains\footnote{Note that in formula (2) of Woodcock's work, there is a plus sign in the term $(z^\ell-z)(w^1_{,x}(x,y)+\varphi^n_x(x,y))$ which is not coherent with the definition ``the shear'' given for the corresponding quantities (denoted $\varphi^n_x$ and $\varphi^n_y$). These quantities are obviously ``the rotations due to shear'', this explains some sign differences between the formulations which appear in the following.}, and the transverse displacement within the layer $\ell$, $w^1(x,y)$ is the transverse displacement at the mid-plane of the first layer. We can see that the transverse displacement is identical for all the layers and does not depend on $z$. However, the superscript $1$ is kept to show that the reference is, in this model, the middle plane of the first layer. Note also that this model needs at this time $4n+1$ unknown functions of the $(x,y)$ coordinates.
\par
The kinematic assumptions of the present formulation are:
\begin{equation}
\label{eq:depl_L}
  \left\{
    \begin{array}{lll}
      u_\alpha(x,y,z) & = & u_\alpha^1(x,y) - z w^1_{,\alpha}(x,y) + \varphi_{\alpha\beta}(z)\gamma^1_{\beta3}(x,y) \\ 
      u_3(x,y,z) & = & w^1(x,y) 
    \end{array} 
  \right.
\end{equation}
The main difference is the lack of the $\ell$ superscript and the consideration of four warping functions $\varphi_{\alpha\beta}(z)$ that will be determined with the use of equations~\eqref{eq:cont_depl_L} and~\eqref{eq:cont_sigm_L}.
\subsubsection{Additional assumptions}
In addition, Woodcock's model (like the second model of reference~\cite{Sun1973}) requires the continuity of in-plane displacements and transverse shear stresses at each of the $n-1$ interfaces:
\begin{subequations}
\begin{empheq}[left=\empheqlbrace]{align}
      u_\alpha^\ell(x,y,\zeta^\ell) & = u_\alpha^{\ell+1}(x,y,\zeta^\ell) \label{eq:cont_depl}\\ 
      \sigma_{\alpha3}^\ell(x,y,\zeta^\ell) & = \sigma_{\alpha3}^{\ell+1}(x,y,\zeta^\ell) \label{eq:cont_sigm}
\end{empheq}
\end{subequations}
where $\sigma_{\alpha3}^\ell$ are the transverse shear stresses in layer $\ell$. These $4\times(n-1)$ equations permit to eliminate all the unknown functions of the upper layers as it is shown in references~\cite{Woodcock2008,Loredo2011,Guyader1978}. This is done by means of transfer matrices and leads to the definition of coefficients $\alpha^{\ell}_{x},\alpha^{\ell}_{xy},\beta^{\ell}_{x}$, etc. As the whole process is quite complicated, it is not reproduced here but presented in details in \ref{sec:LayerCoeffs}. The model finally reduces to $4n+1-(4n-4)=5$ unknown functions of the first layer: two in-plane displacements $u_\alpha^1(x,y,z^1)$, a transverse displacement $w^1(x,y)$, and two transverse shear strains $\gamma^1_{\alpha3}(x,y)$. These generalized displacements differ from the Mindlin--Reissner plate model in which rotations $\phi_\alpha^1(x,y) = w^1_{,\alpha}(x,y)-\gamma^1_{\alpha3}(x,y)$ are considered. As previously mentioned, these quantities are those of the first layer, whereas in plate models, they are generally defined at the mid-plane of the laminate.
\par
For the present formulation, which is already written in function of the above five unknown functions, the same physical assumptions are done. They are written in a slightly different manner:
\begin{subequations}
\begin{empheq}[left=\empheqlbrace]{align}
      \lim_{z \rightarrow \zeta^\ell-} u_\alpha(x,y,z) & = \lim_{z \rightarrow \zeta^\ell+} u_\alpha(x,y,z) \label{eq:cont_depl_L}\\ 
      \lim_{z \rightarrow \zeta^\ell-} \sigma_{\alpha3}(x,y,z) & = \lim_{z \rightarrow \zeta^\ell+}\sigma_{\alpha3}(x,y,z) \label{eq:cont_sigm_L}
\end{empheq}
\end{subequations}
The advantage of this formulation is that, as shown below, the four warping functions $\varphi_{\alpha\beta}(z)$ are easily obtained by means of the above conditions, making the displacement field explicit.
\subsubsection{Final form}
As told before, the elimination of the displacements of the upper layers in Woodcock's formulation is a quite complicated process and does not lead to an explicit form for the displacement field. The displacement field is then written independently for each layer with the help of coefficients that must be computed for each layer.
\par 
For the present model, the building of the warping functions with the help of equations~\eqref{eq:cont_depl_L} and~\eqref{eq:cont_sigm_L} is presented below. First, let us compute the strain field from equations~\eqref{eq:depl_L}:
\begin{subequations}\label{eq:strains}
\begin{empheq}[left=\empheqlbrace]{align}
  \epsilon_{\alpha\beta}(x,y,z) & = \epsilon_{\alpha\beta}^1(x,y) - z w^1_{,\alpha\beta}(x,y) 
	       + \frac{1}{2}\left(\varphi_{\alpha\gamma}(z)\gamma^1_{\gamma3,\beta}(x,y)+\varphi_{\beta\gamma}(z)\gamma^1_{\gamma3,\alpha}(x,y)\right) \label{eq:in_plane_strains}\\ 
  \epsilon_{\alpha3}(x,y,z) & = \frac{1}{2}\varphi'_{\alpha\beta}(z)\gamma^1_{\beta3}(x,y) \label{eq:transverse_strains}\\ 
  \epsilon_{33}(x,y,z) & = 0 \phantom{\frac{1}{2}}
\end{empheq}
\end{subequations}
Note that $\epsilon_{\alpha\beta}^1$, $w^1_{,\alpha\beta}$, $\gamma^1_{\alpha3,\beta}$ and $\gamma^1_{\alpha3}$ form a set of $12$ independent generalized strains to consider in this model.
\par
With the help of equation~\eqref{eq:transverse_strains} and Hooke's law, the transverse shear stresses can be formulated:
%
\begin{equation}
  \sigma_{\alpha3}(x,y,z) = 2C_{\alpha3\beta3}\epsilon_{\beta3} = C_{\alpha3\beta3}(z)\varphi'_{\beta\gamma}(z)\gamma^1_{\gamma3}(x,y) \label{eq:transverse_stresses1}
\end{equation}
%
On the above formula, the 2 factor comes from the symmetry of the stiffness and strain tensors:
\begin{align}
	\sigma_{\alpha3}=C_{\alpha3\beta3}\epsilon_{\beta3}+C_{\alpha33\beta}\epsilon_{3\beta}=2C_{\alpha3\beta3}\epsilon_{\beta3}
\end{align}
The additional conditions~\eqref{eq:cont_depl_L} and~\eqref{eq:cont_sigm_L} can now be written in terms of conditions on the warping functions:
\begin{subequations}
\begin{empheq}[left=\empheqlbrace]{align}
      &\lim_{z \rightarrow \zeta^\ell-} \varphi_{\alpha\beta}(z)  = \lim_{z \rightarrow \zeta^\ell+} \varphi_{\alpha\beta}(z) \label{eq:cont_depl_L2}\\ 
      &\lim_{z \rightarrow \zeta^\ell-} C_{\alpha3\beta3}(z)\varphi'_{\beta\gamma}(z)  = \lim_{z \rightarrow \zeta^\ell+} C_{\alpha3\beta3}(z)\varphi'_{\beta\gamma}(z) \label{eq:cont_sigm_L2}
\end{empheq}
\end{subequations}
Equations~\eqref{eq:cont_depl_L2} represent the continuity of the warping functions at the $n-1$ interfaces whereas equations~\eqref{eq:cont_sigm_L2} represent ``jump'' conditions on the derivatives. For a laminate, the simplest way to choose the warping functions is to take their derivatives as step functions, then, the warping functions are continuous piecewise linear functions of $z$. For the first layer \emph{i. e.} for $z\in[\zeta^0,\zeta^1]$, the choice of $\varphi'_{\alpha\beta}(z)=\delta^K_{\alpha\beta}$ is done, where $\delta^K_{\alpha\beta}$ is the Kronecker symbol. This choice corresponds to Woodcock's model.
\par
This leads to the following expression for the derivatives of the warping functions:
\begin{equation}
  \varphi'_{\alpha\beta}(z) = 4S_{\alpha3\gamma3}(z) C_{\gamma3\delta3}(z^1)\delta_{\delta\beta}=4S_{\alpha3\gamma3}(z) C_{\gamma3\beta3}(z^1) \label{eq:def_deriv_warp}
\end{equation}
The warping functions are obtained integrating \eqref{eq:def_deriv_warp}. Integration constants are chosen to give $\varphi_{\alpha\beta}(0)=0$. That means that the transverse shear strains have no effect on the displacement of the middle plane of the first layer, which is normal because it is precisely the reference plane for the model. Hence the warping functions could be written:
\begin{equation}\label{eq:warping_functions}
  \varphi_{\alpha\beta}(z) = 4C_{\gamma3\beta3}(z^1)\int_0^z S_{\alpha3\gamma3}(\zeta)\text{d}\zeta
\end{equation}
\subsection{Stresses}
The strains obtained in formula~\eqref{eq:strains} permit to compute the stresses:
\begin{subequations}\label{eq:stresses}
\begin{empheq}[left=\empheqlbrace]{align}
  \sigma_{\alpha\beta}(x,y,z) & = Q_{\alpha\beta\gamma\delta}(z)\left(\epsilon_{\gamma\delta}^1(x,y) - z w^1_{,\gamma\delta}(x,y)
	       + \varphi_{\gamma\mu}(z)\gamma^1_{\mu3,\delta}(x,y)\right) \label{eq:in_plane_stresses}\\ 
  \sigma_{\alpha3}(x,y,z) & = C_{\alpha3\beta3}\varphi'_{\beta\mu}(z)\gamma^1_{\mu3}(x,y) \label{eq:transverse_stresses2}\\ 
  \sigma_{33}(x,y,z) & = 0
\end{empheq}
\end{subequations} 
where $Q_{\alpha\beta\gamma\delta}$ are the reduced generalized plane strain stiffnesses. The vanishing of the term $\tfrac{1}{2}$ of equation~\eqref{eq:in_plane_strains} into equation~\eqref{eq:in_plane_stresses} is not evident. Let us demonstrate it, omitting the $x$, $y$, and $z$ coordinates:
\begin{align}
  \nonumber
	\frac{1}{2}Q_{\alpha\beta\gamma\delta}\left(\varphi_{\gamma\mu}\gamma^1_{\mu3,\delta}+\varphi_{\delta\mu}\gamma^1_{\mu3,\gamma}\right)
	&= \frac{1}{2}Q_{\alpha\beta\gamma\delta}\varphi_{\gamma\mu}\gamma^1_{\mu3,\delta}+\frac{1}{2}Q_{\alpha\beta\gamma\delta}\varphi_{\delta\mu}\gamma^1_{\mu3,\gamma} \\
	&= \frac{1}{2}Q_{\alpha\beta\gamma\delta}\varphi_{\gamma\mu}\gamma^1_{\mu3,\delta}+\frac{1}{2}Q_{\alpha\beta\delta\gamma}\varphi_{\gamma\mu}\gamma^1_{\mu3,\delta} \\
	\nonumber
	&= Q_{\alpha\beta\gamma\delta}\varphi_{\gamma\mu}\gamma^1_{\mu3,\delta}\phantom{\frac{1}{2}}
\end{align}
\subsection{Strain energy}\label{sec:strain_energy}
It is possible to compute the strain energy density $\delta J=1/2\epsilon_{ij}\sigma_{ij}$ from formulas~\eqref{eq:strains} and~\eqref{eq:stresses} and integrate it over the thickness to obtain a strain energy surface density $J(x,y)$. This is the method followed by Woodcock and other previous authors. Woodcock has found an expression which is the sum of $69$ terms. In fact there are only $55$ independent terms as we shall see below. 
\par
In this paper, this method was also reproduced to find Woodcock's coefficients given in \ref{sec:EnergyCoeffs} but in the following lines, a different approach is presented. Let us start with the strain energy surface density:
\begin{align}\label{eq:StrainEnergy1}
  J&=\frac{1}{2}\int_{\zeta_0}^{\zeta_n}\epsilon_{ij}\sigma_{ij}\text{d}z
	  =\frac{1}{2}\int_{\zeta_0}^{\zeta_n}\left(\epsilon_{\alpha\beta}\sigma_{\alpha\beta}+2\epsilon_{\alpha3}\sigma_{\alpha3}+\epsilon_{33}\sigma_{33}\right)\text{d}z \nonumber \\
	 &=\frac{1}{2}\int_{\zeta_0}^{\zeta_n}
	  \left[\left(\epsilon_{\alpha\beta}^1 - z w^1_{,\alpha\beta} + \frac{1}{2}\left(\varphi_{\alpha\gamma}(z)\gamma^1_{\gamma3,\beta}
		  +\varphi_{\beta\gamma}(z)\gamma^1_{\gamma3,\alpha}\right)\right)\sigma_{\alpha\beta}
			+2\frac{1}{2}\varphi'_{\alpha\beta}(z)\gamma^1_{\beta3}\sigma_{\alpha3}\right]\text{d}z \nonumber \\
	 &=\frac{1}{2}\int_{\zeta_0}^{\zeta_n}
	  \left[\left(\epsilon_{\alpha\beta}^1 - z w^1_{,\alpha\beta} + \varphi_{\alpha\gamma}(z)\gamma^1_{\gamma3,\beta}\right)\sigma_{\alpha\beta}
		  +\varphi'_{\alpha\beta}(z)\gamma^1_{\beta3}\sigma_{\alpha3}\right]\text{d}z
\end{align}
%
%
It can also be written
\begin{align}\label{eq:StrainEnergy2}
  J =\frac{1}{2}
	  \left[\epsilon_{\alpha\beta}^1 N_{\alpha\beta} - w^1_{,\alpha\beta} M_{\alpha\beta} + \gamma^1_{\gamma3,\beta} P_{\gamma\beta} + \gamma^1_{\beta3} Q_{\beta} \right]
\end{align}
naturally introducing the following quantities which are the generalized forces,
\begin{subequations}\label{eq:generalized_forces1}
\begin{empheq}[left=\empheqlbrace]{align}
  \{N_{\alpha\beta},M_{\alpha\beta},P_{\gamma\beta}\}&=\int^{\zeta^n}_{\zeta^0} \{1,z,\varphi_{\alpha\gamma}(z)\} \sigma_{\alpha\beta}(z) \text{d}z \\
  Q_{\beta}&=\int^{\zeta^n}_{\zeta^0} \varphi'_{\alpha\beta}(z) \sigma_{\alpha3}(z) \text{d}z
\end{empheq}
\end{subequations}
each associated with a corresponding generalized displacement in the strain energy formula~\eqref{eq:StrainEnergy2}. 
\par
$N_{\alpha\beta}$ and $M_{\alpha\beta}$ are respectively the classical plate membrane forces and bending moments, and $P_{\alpha\beta}$ and $Q_{\alpha}$ are special moments associated with the warping functions, \emph{i. e.} associated with the transverse shear behavior. Note that $P_{\alpha\beta} \ne P_{\beta\alpha}$ in the general case, leading to a set of $12$ generalized forces.
\par
Let us calculate these generalized forces with the help of equations~\eqref{eq:stresses} and~\eqref{eq:generalized_forces1}. They are (some explanations are given below): 
\begin{subequations}\label{eq:generalized_forces2}
\begin{empheq}[left=\empheqlbrace]{align}
  N_{\alpha\beta} &=	A_{\alpha\beta\gamma\delta}\epsilon_{\gamma\delta}^1 + B_{\alpha\beta\gamma\delta} (-w^1_{,\gamma\delta})+ E_{\alpha\beta\mu\delta} \gamma^1_{\mu3,\delta}\\
	M_{\alpha\beta} &=	B_{\alpha\beta\gamma\delta}\epsilon_{\gamma\delta}^1 + D_{\alpha\beta\gamma\delta} (-w^1_{,\gamma\delta})+ F_{\alpha\beta\mu\delta} \gamma^1_{\mu3,\delta}\\
	P_{\alpha\beta} &=	E_{\gamma\delta\alpha\beta}\epsilon_{\gamma\delta}^1 + F_{\gamma\delta\alpha\beta} (-w^1_{,\gamma\delta})+ G_{\alpha\beta\mu\delta} \gamma^1_{\mu3,\delta}\\
	Q_{\alpha} &=	H_{\alpha3\beta3}\gamma^1_{\beta3}
\end{empheq}
\end{subequations}
where the following generalized stiffnesses have been introduced:
\begin{subequations}\label{eq:generalized stiffnesses1}
\begin{empheq}[left=\empheqlbrace]{align}
  \{A_{\alpha\beta\gamma\delta},B_{\alpha\beta\gamma\delta},D_{\alpha\beta\gamma\delta},E_{\alpha\beta\mu\delta},F_{\alpha\beta\mu\delta},G_{\nu\beta\mu\delta}\}&=\int^{\zeta^n}_{\zeta^0} Q_{\alpha\beta\gamma\delta}\{1,z,z^2,\varphi_{\gamma\mu}(z),z\varphi_{\gamma\mu}(z),\varphi_{\alpha\nu}(z)\varphi_{\gamma\mu}(z)\}\text{d}z \\
  H_{\alpha3\beta3}&=\int^{\zeta^n}_{\zeta^0} \varphi'_{\gamma\alpha}(z) C_{\gamma3\delta3} \varphi'_{\delta\beta}(z) \text{d}z
\end{empheq}
\end{subequations}
The $N_{\alpha\beta}$ and $M_{\alpha\beta}$ computation is straightforward but special attention must be paid to the calculation of the $P_{\alpha\beta}$, revealing some uncommon symmetries:
\begin{align}
	P_{\alpha\beta} 
	& =	\int^{\zeta^n}_{\zeta^0} \varphi_{\mu\alpha}(z) \sigma_{\mu\beta}(z) \text{d}z \nonumber \\
	& = \int^{\zeta^n}_{\zeta^0} \varphi_{\mu\alpha}(z) Q_{\mu\beta\gamma\delta}(z) \left(\epsilon_{\gamma\delta}^1 - z w^1_{,\gamma\delta} + \varphi_{\gamma\nu}(z)\gamma^1_{\nu3,\delta}\right) \text{d}z \nonumber \\
	& = E_{\gamma\delta\alpha\beta}\epsilon_{\gamma\delta}^1 + F_{\gamma\delta\alpha\beta} (-w^1_{,\gamma\delta})+ G_{\alpha\beta\mu\delta} \gamma^1_{\mu3,\delta} \phantom{\int^{\zeta^n}_{\zeta^0} }
\end{align}
In this last expression, the $E_{\gamma\delta\alpha\beta}$ and $F_{\gamma\delta\alpha\beta}$ are identified with the help of the major symmetry of the $Q_{\mu\beta\gamma\delta}(z)$ tensor.
\par 
The $\mathbf{A}$, $\mathbf{B}$ and $\mathbf{D}$ tensors inherit the symmetries of Hooke's tensor, a symmetry for each pair of indices called the \emph{minor symmetries}, and the \emph{major symmetry} which permits the swap of the two pairs of indices, this last one being related to the existence of a strain energy. The $\mathbf{E}$ and $\mathbf{F}$ tensors loose the symmetry on the last pair of indices, forcing the major symmetry to disappear. The $\mathbf{G}$ tensor looses the symmetry on the two pairs of indices, but keeps the major symmetry:
%
\begin{empheq}[left=\empheqlbrace]{alignat=2}\label{eq:tensors_symmetries}
  \nonumber 
	&\text{for the }\mathbf{A}\text{, }\mathbf{B}\text{, }\mathbf{D} \text{ tensors:} &\quad
	  &A_{\beta\alpha\gamma\delta}   =  A_{\alpha\beta\gamma\delta}   =  A_{\gamma\delta\alpha\beta}   =  A_{\gamma\delta\beta\alpha}  \\
  &\text{for the }\mathbf{E}\text{, }\mathbf{F}\text{ tensors:} &\quad
    &E_{\beta\alpha\gamma\delta}   =  E_{\alpha\beta\gamma\delta} \neq E_{\gamma\delta\alpha\beta} \neq E_{\gamma\delta\beta\alpha}  \\
  \nonumber 
	&\text{for the }\mathbf{G}\text{ tensor:} &\quad
    &G_{\beta\alpha\gamma\delta} \neq G_{\alpha\beta\gamma\delta}   =  G_{\gamma\delta\alpha\beta} \neq G_{\delta\gamma\beta\alpha} 
\end{empheq}
%
Hence there are $6$ independent components for $\mathbf{A}$, $\mathbf{B}$ and $\mathbf{D}$, $12$ for $\mathbf{E}$ and $\mathbf{F}$, $10$ for $\mathbf{G}$, and $3$ for $\mathbf{H}$. So this plate model has a total of $55$ independent stiffness coefficients in the most general case.
\subsection{Static laminate behavior}
The generalized forces are set, by type, into vectors:
\begin{equation}\label{eq:generalized_forces}
  \mathbf{N}=
  \begin{Bmatrix}
    N_{11} \\
		N_{22} \\
		N_{12}
  \end{Bmatrix}
  \quad
  \mathbf{M}=
  \begin{Bmatrix}
    M_{11} \\
		M_{22} \\
	  M_{12}
  \end{Bmatrix}
  \quad
  \mathbf{P}=
  \begin{Bmatrix}
    P_{11} \\
		P_{22} \\
		P_{12} \\
	  P_{21}
  \end{Bmatrix}
	\quad
  \mathbf{Q}=
  \begin{Bmatrix}
    Q_{1} \\
		Q_{2}
  \end{Bmatrix}
\end{equation} 
and the same is done for the corresponding generalized strains:
\begin{equation}\label{eq:generalized_strains}
  \boldsymbol{\epsilon}=
  \begin{Bmatrix}
    \epsilon^1_{11} \\
		\epsilon^1_{22} \\
		2\epsilon^1_{12}
  \end{Bmatrix}
  \quad
  \boldsymbol{\kappa}=
  \begin{Bmatrix}
    -w^1_{,11} \\
		-w^1_{,22} \\
	  -2w^1_{,12}
  \end{Bmatrix}
  \quad
  \mathbf{\Gamma}=
  \begin{Bmatrix}
    \gamma^1_{13,1} \\
		\gamma^1_{23,2} \\
		\gamma^1_{13,2} \\
	  \gamma^1_{23,1}
  \end{Bmatrix}
  \quad
  \boldsymbol{\gamma}=
  \begin{Bmatrix}
    \gamma^1_{13} \\
		\gamma^1_{23}
  \end{Bmatrix}
\end{equation} 
Hence, generalized forces are linked with the generalized strains by the $10\times10$ and $2\times2$ following stiffness matrices:
\begin{equation}\label{eq:behavior}
\begin{Bmatrix}
  \mathbf{N} \\
	\mathbf{M} \\
	\mathbf{P}
\end{Bmatrix}
=
\begin{bmatrix}
  \mathbf{A} & \mathbf{B} & \mathbf{E} \\
	\mathbf{B} & \mathbf{D} & \mathbf{F} \\
	\mathbf{E^T} & \mathbf{F^T} & \mathbf{G}
\end{bmatrix}
\begin{Bmatrix}
  \boldsymbol{\epsilon} \\
	\boldsymbol{\kappa} \\
	\boldsymbol{\Gamma}
\end{Bmatrix}
\quad
\begin{Bmatrix}
  \mathbf{Q}
\end{Bmatrix}
=
\begin{bmatrix}
  \mathbf{H}
\end{bmatrix}
\begin{Bmatrix}
	\boldsymbol{\gamma}
\end{Bmatrix}
\end{equation} 
\subsection{Kinetic energy}
The kinetic energy surface density $E_c(x,y)$ of the structure is:
\begin{align}\label{eq:KineticEnergy1}
  E_c(x,y)&=\frac{1}{2}\int_{\zeta^0}^{\zeta^n}
	        \rho(x,y,z)\dot{u}_{i}(x,y,z)\dot{u}_{i}(x,y,z)\text{d}z \nonumber \\
	        &=\frac{1}{2}\int_{\zeta^0}^{\zeta^n}
	  \rho(z)\bigg[\left(\dot{u}_\alpha^1 - z \dot{w}^1_{,\alpha} + \varphi_{\alpha\beta}(z)\dot{\gamma}^1_{\beta3}\right)
		      \left(\dot{u}_\alpha^1 - z \dot{w}^1_{,\alpha} + \varphi_{\alpha\beta}(z)\dot{\gamma}^1_{\beta3}\right) + (\dot{w}^1)^2\bigg]\text{d}z \nonumber \\
	        &=\frac{1}{2}\int_{\zeta^0}^{\zeta^n}
	  \rho(z)\bigg[\dot{u}_\alpha^1\dot{u}_\alpha^1 - 2 z \dot{u}_\alpha^1 \dot{w}^1_{,\alpha} + 2 \dot{u}_\alpha^1 \varphi_{\alpha\beta}(z)\dot{\gamma}^1_{\beta3} + z^2 \dot{w}^1_{,\alpha}\dot{w}^1_{,\alpha} \nonumber \\
		      &\phantom{=\frac{1}{2}\int_{\zeta^0}^{\zeta^n}\rho(z)\bigg[}- 2 z \dot{w}^1_{,\alpha} \varphi_{\alpha\beta}(z)\dot{\gamma}^1_{\beta3} + \varphi_{\alpha\beta}(z)\dot{\gamma}^1_{\beta3} \varphi_{\alpha\mu}(z)\dot{\gamma}^1_{\mu3} + (\dot{w}^1)^2\bigg]\text{d}z
\end{align}
for concision $x$ and $y$ have been omitted for the last two lines of this formula.
\par
Let us now introduce the following generalized mass:
\begin{equation}\label{eq:generalized_mass}
  \{R,S,T,U_{\alpha\beta},V_{\alpha\beta},W_{\alpha\beta}\} = \int^{\zeta^n}_{\zeta^0} \rho(z)\{1,z,z^2,\varphi_{\alpha\beta}(z),\varphi_{\alpha\beta}(z)z,\varphi_{\mu\alpha}(z)\varphi_{\mu\beta}(z)\} \text{d}z
\end{equation} 
Note that the $U_{\alpha\beta}$ and $V_{\alpha\beta}$ are antisymmetric tensors but $W_{\alpha\beta}$ is symmetric. Then, there are $14$ independent mass coefficients to consider. The kinetic energy surface density can now be written:
\begin{align}\label{eq:KineticEnergy2}
  E_c(x,y)&=\frac{1}{2}
	  \bigg(R \dot{u}_\alpha^1 \dot{u}_\alpha^1 - 2 S \dot{u}_\alpha^1 \dot{w}^1_{,\alpha} + 2 U_{\alpha\beta} \dot{u}_\alpha^1 \dot{\gamma}^1_{\beta3} + T \dot{w}^1_{,\alpha}\dot{w}^1_{,\alpha} \nonumber \\
		      &\phantom{\bigg(}- 2 V_{\alpha\beta} \dot{w}^1_{,\alpha} \dot{\gamma}^1_{\beta3} + W_{\alpha\beta} \dot{\gamma}^1_{\alpha3} \dot{\gamma}^1_{\beta3} + R(\dot{w}^1)^2\bigg)\text{d}z
\end{align}
\subsection{Laminate equations of motion}
Let us recall the equilibrium conditions within a solid. Without loss of generality, body forces are neglected here, and the previous convention on indices is kept:
\begin{subequations}\label{eq:equilibrium0}
\begin{empheq}[left=\empheqlbrace]{align}
  \sigma_{\alpha\beta,\beta}+\sigma_{\alpha3,3}=\rho \ddot{u}_{\alpha} \label{eq:equilibrium0_membrane} \\
  \sigma_{\alpha3,\alpha}+\sigma_{33,3}=\rho \ddot{u}_{3}
\end{empheq}
\end{subequations} 
\par
Integrating the equations of equilibrium~\eqref{eq:equilibrium0} over the thickness with the help of formulas~\eqref{eq:depl_L}, \eqref{eq:generalized_forces1} and~\eqref{eq:generalized_mass} leads to:
%
%
\begin{subequations}\label{eq:equilibrium1}
\begin{empheq}[left=\empheqlbrace]{align}
  N_{\alpha\beta,\beta}+[\sigma_{\alpha3}(z)]^{\zeta^n}_{\zeta^0}&=R \ddot{u}^1_{\alpha} - S \ddot{w}^1_{,\alpha}+U_{\alpha\beta}\ddot{\gamma}^1_{\beta3} \\
  Q^c_{\alpha,\alpha}+[\sigma_{33}(z)]^{\zeta^n}_{\zeta^0}&=R \ddot{w}^1 \label{eq:equilibrium1b}
\end{empheq}
\end{subequations} 
where the $Q^c_{\alpha}$ are the classical shear forces. In order to get more equations, weighted integrals over the thickness of equation~\eqref{eq:equilibrium0_membrane} are computed. Weight functions are $z$ and $\varphi_{\alpha\gamma}(z)$. It gives four more equations:%
%
%
\begin{subequations}\label{eq:equilibrium2}
\begin{empheq}[left=\empheqlbrace]{align}
 & M_{\alpha\beta,\beta}+[\sigma_{\alpha3}(z)z]^{\zeta^n}_{\zeta^0}-Q^c_{\alpha} = S \ddot{u}^1_{\alpha} - T \ddot{w}^1_{,\alpha}+V_{\alpha\beta}\ddot{\gamma}^1_{\beta3} \label{eq:equilibrium2a}\\
 & P_{\gamma\beta,\beta}+[\varphi_{\alpha\gamma}(z)\sigma_{\alpha3}(z)]^{\zeta^n}_{\zeta^0}-Q_{\gamma} = U_{\alpha\gamma} \ddot{u}^1_{\alpha} - V_{\alpha\gamma} \ddot{w}^1_{,\alpha}+W_{\gamma\beta}\ddot{\gamma}^1_{\beta3} 
\end{empheq}
\end{subequations} 
Let $q=[\sigma_{33}(z)]^{\zeta^n}_{\zeta^0}$ denote the value of the transverse loading and suppose there is no tangential forces on the top and bottom of the plate, so $\sigma_{\alpha3}(-h/2)=\sigma_{\alpha3}(h/2)=0$. We shall note that there is no generalized strains corresponding to the classical shear forces $Q^c_{\alpha}$. They must be eliminated. It is done by setting values of $Q^c_{\alpha}$ obtained from formula~\eqref{eq:equilibrium2a} into equation~\eqref{eq:equilibrium1b}. This leads to the plate equilibrium system of equations:
\begin{subequations}\label{eq:equilibrium3}
\begin{empheq}[left=\empheqlbrace]{align}
  &N_{\alpha\beta,\beta}=R \ddot{u}^1_{\alpha} - S \ddot{w}^1_{,\alpha}+U_{\alpha\beta}\ddot{\gamma}^1_{\beta3} \\
  &M_{\alpha\beta,\beta\alpha} + q =R \ddot{w}^1 + S \ddot{u}^1_{\alpha,\alpha} - T \ddot{w}^1_{,\alpha\alpha}+V_{\alpha\beta}\ddot{\gamma}^1_{\beta3,\alpha} \\
  &P_{\alpha\beta,\beta}-Q_{\alpha}=U_{\beta\alpha} \ddot{u}^1_{\beta} - V_{\beta\alpha} \ddot{w}^1_{,\beta}+W_{\alpha\beta}\ddot{\gamma}^1_{\beta3} 
\end{empheq}
\end{subequations} 
\subsection{Link with Woodcock's model}
\subsubsection{Strain energy}
The behavior matrices of equation~\eqref{eq:behavior} can be formulated using Woodcock's 69 $\lambda_i$ energy coefficients. Computation of these coefficients is detailed in \ref{sec:LayerCoeffs} and \ref{sec:EnergyCoeffs} and the proof of the link with stiffnesses of formulas~\ref{eq:generalized stiffnesses1} is given in \ref{sec:Proof}. The two matrices of equation~\eqref{eq:behavior} can be written:
\begin{gather}\label{eq:behaviorWoodcock}
	\begin{bmatrix}
		\mathbf{A} & \mathbf{B} & \mathbf{E} \\
		\mathbf{B} & \mathbf{D} & \mathbf{F} \\
		\mathbf{E^T} & \mathbf{F^T} & \mathbf{G}
	\end{bmatrix}
	=
  \frac{1}{4}\left[\begin{array}{ccc|ccc|cccc}
		   4\lambda_{3}  &  2\lambda_{21} &  2\lambda_{47} & -2\lambda_{5}  & -2\lambda_{15} & -\lambda_{45} & -2\lambda_{6}  & -2\lambda_{20} & -2\lambda_{46} & -2\lambda_{52} \\
		   2\lambda_{21} &  4\lambda_{9}  &  2\lambda_{62} & -2\lambda_{15} & -2\lambda_{11} & -\lambda_{60} & -2\lambda_{18} & -2\lambda_{12} & -2\lambda_{61} & -2\lambda_{67} \\
		   2\lambda_{47} &  2\lambda_{62} &  4\lambda_{25} & -2\lambda_{41} & -2\lambda_{56} & -\lambda_{29} & -2\lambda_{44} & -2\lambda_{59} & -2\lambda_{32} & -2\lambda_{34} \\
    \hline
		  -2\lambda_{5}  & -2\lambda_{15} & -2\lambda_{41} &  4\lambda_{1}  &  2\lambda_{13} &  \lambda_{39} &  2\lambda_{4}  &  2\lambda_{14} &  2\lambda_{40} &  2\lambda_{48} \\
		  -2\lambda_{15} & -2\lambda_{11} & -2\lambda_{56} &  2\lambda_{13} &  4\lambda_{7}  &  \lambda_{54} &  2\lambda_{16} &  2\lambda_{10} &  2\lambda_{55} &  2\lambda_{63} \\
		  - \lambda_{45} & - \lambda_{60} & - \lambda_{29} &   \lambda_{39} &   \lambda_{54} &  \lambda_{22} &   \lambda_{42} &   \lambda_{57} &   \lambda_{27} &   \lambda_{28} \\
    \hline
		  -2\lambda_{6}  & -2\lambda_{18} & -2\lambda_{44} &  2\lambda_{4}  &  2\lambda_{16} &  \lambda_{42} &  4\lambda_{2}  &  2\lambda_{17} &  2\lambda_{43} &  2\lambda_{50} \\
		  -2\lambda_{20} & -2\lambda_{12} & -2\lambda_{59} &  2\lambda_{14} &  2\lambda_{10} &  \lambda_{57} &  2\lambda_{17} &  4\lambda_{8}  &  2\lambda_{58} &  2\lambda_{65} \\
   	  -2\lambda_{46} & -2\lambda_{61} & -2\lambda_{32} &  2\lambda_{40} &  2\lambda_{55} &  \lambda_{27} &  2\lambda_{43} &  2\lambda_{58} &  4\lambda_{23} &  2\lambda_{31} \\
		  -2\lambda_{52} & -2\lambda_{67} & -2\lambda_{34} &  2\lambda_{48} &  2\lambda_{63} &  \lambda_{28} &  2\lambda_{50} &  2\lambda_{65} &  2\lambda_{31} &  4\lambda_{24}  	
		\end{array}\right]
\end{gather}
and:
\begin{gather}
	 \begin{bmatrix}
		\mathbf{H}
	\end{bmatrix}
	=
  \frac{1}{4}\left[\begin{array}{cc}
		4\lambda_{37} & 2\lambda_{69} \\
		2\lambda_{69} & 4\lambda_{38} 
	\end{array}\right]
\end{gather}
with:
\begin{gather}
	\nonumber\lambda_{15}=\lambda_{19}\\
	\nonumber\lambda_{25}=\lambda_{26}=\lambda_{36}/2\\
	\nonumber\lambda_{29}=\lambda_{30}\\
	\nonumber\lambda_{32}=\lambda_{33}\\
	\nonumber\lambda_{34}=\lambda_{35}\\
	\lambda_{41}=\lambda_{49}=\lambda_{45}/2 \label{eq:equallambdas} \\
	\nonumber\lambda_{44}=\lambda_{51}\\
	\nonumber\lambda_{47}=\lambda_{53}\\
	\nonumber\lambda_{56}=\lambda_{64}=\lambda_{60}/2\\
	\nonumber\lambda_{59}=\lambda_{66}\\
	\nonumber\lambda_{62}=\lambda_{68}
\end{gather}
These above $14$ equalities show that only $69-14=55$ strain energy coefficients are independent, which is the number previously announced after the examination of symmetries of tensors done in section~\ref{sec:strain_energy} starting from formula~\eqref{eq:tensors_symmetries}. 
\par
The presence of minus sign in the blocs corresponding to $\mathbf{B}$ and $\mathbf{E}$ blocs is not easy to understand because it is due to two sign differences between the two formulations: $-w_{,\alpha\beta}$ are used here instead of $w_{,\alpha\beta}$ in Woodcock's work, and $\gamma_{\alpha3}$ are used here instead of $\varphi_{\alpha}=-\gamma_{\alpha3}$ (hence $\varphi_{\alpha,\beta}=-\gamma_{\alpha3,\beta}$) in Woodcock's work. The combination of these two sign changes gives an explanation to this point.
\subsubsection{Kinetic energy}
Let us put the generalized speeds into a vector:
\begin{align}\label{eq:SpeedVector}
  \boldsymbol{\Omega}^T=\begin{Bmatrix} \dot{u}_1^1 & \dot{u}_2^1 & \dot{w}^1 & -\dot{w}^1_{,1} & -\dot{w}^1_{,2} & \dot{\gamma}^1_{13} & \dot{\gamma}^1_{23} \end{Bmatrix}
\end{align}
which permits to define a generalized mass matrix $\boldsymbol{\Xi}$:
\begin{gather}\label{eq:KineticEnergy3}
  E_c(x,y)=\frac{1}{2} \boldsymbol{\Omega}^T \boldsymbol{\Xi} \boldsymbol{\Omega}
\end{gather}
Identifying equation~\eqref{eq:KineticEnergy2} with formula~(18) of Woodcock's work\footnote{The terms $\delta_i$ for $14\le i\le18$ are missing in Woodcock's formula (18) and also in the list of $\delta_i$ he produced. They are introduced here, but the corrected formula (18) is not given: one can easily obtain it with the present mass matrix or following Woodcock's procedure.} shows that the expression of the generalized mass matrix can be put in the two alternative forms:
\begin{gather}\label{eq:MassMatrix}
  \boldsymbol{\Xi}
	=
	\left[\begin{array}{ccc|cc|cc}
		   R &  0 &  0 &  S &  0 & U_{11} & U_{12} \\
		   0 &  R  & 0 &  0 &  S & U_{21} & U_{22} \\
		   0 &  0 &  R &  0 &  0 & 0 & 0\\
    \hline
		   S &  0 &  0 & T &  0 & V_{11} & V_{12} \\
		   0 &  S  & 0 &  0 & T & V_{21} & V_{22} \\
    \hline
		   U_{11} &  U_{21} &  0 & V_{11} & V_{21} & W_{11} &  W_{12} \\
		   U_{12} &  U_{22}  & 0 & V_{12} & V_{22} & W_{12} &  W_{22}
		\end{array}\right]
		=
	\frac{1}{2}
	\left[\begin{array}{ccc|cc|cc}
	  2\delta_{3}  &      0       &      0       & -\delta_{5} &      0       & -\delta_{6} & -\delta_{18} \\
	       0       & 2\delta_{9}  &      0       &      0      & -\delta_{11} & -\delta_{17}& -\delta_{12} \\
         0       &      0       & 2\delta_{13} &      0      &      0       &      0      &      0       \\
    \hline
		-\delta_{5}  &      0       &      0       & 2\delta_{1} &      0       &  \delta_{4} &  \delta_{15} \\
		     0       & -\delta_{11} &      0       &      0      & 2\delta_{7}  &  \delta_{16}&  \delta_{10} \\
    \hline
		-\delta_{6}  & -\delta_{17} &      0       &  \delta_{4} &  \delta_{16} & 2\delta_{2} &  \delta_{14} \\
		-\delta_{18} & -\delta_{12} &      0       &  \delta_{15}&  \delta_{10} &  \delta_{14}& 2\delta_{8}
		\end{array}\right]
\end{gather}
Note that:
\begin{gather}
	\nonumber\delta_{3}=\delta_{9}=\delta_{13} \\
	\delta_{5}=\delta_{11} \label{eq:equaldeltas} \\
	\nonumber\delta_{1}=\delta_{7}
\end{gather}
These above $4$ equalities show that only $13-4=9$ coefficients were independent in Woodcock's work (that can be verified in Appendix A. of reference~\cite{Woodcock2008}), but with the $5$ missing coefficients, this leads to the announced $14$ independent generalized mass coefficients. Complete proof of this relationship, including the missing coefficients, is given in \ref{sec:Proof}. 

%% file: 03_PlaceOfModel.tex
\section{Place of this model relatively to other plate models}
\subsection{Discussion based on the formulations}\label{sec:otherformulations}
The present model is more general than the Love--Kirchhoff, Mindlin--Reissner and Reddy's models because it integrates additional information on the laminate behavior. Setting $\varphi_{\alpha\beta}(z)=0$ in the present model gives the Love--Kirchhoff model. Matrices $\mathbf{E}$, $\mathbf{F}$, $\mathbf{G}$ and $\mathbf{H}$ are null, so the system~\eqref{eq:behavior} can be reduced to the classical $6\times6$ four-block membrane-bending behavior matrix. The shear matrix $\mathbf{H}$ is null. Mass terms $U_{\alpha\beta}$, $V_{\alpha\beta}$, $W_{\alpha\beta}$ are also null, so the motion equations reduces to three equations:
\begin{subequations}\label{eq:equilibriumLK}
\begin{empheq}[left=\empheqlbrace]{align}
  &N_{\alpha\beta,\beta}=R \ddot{u}^1_{\alpha} - S \ddot{w}^1_{,\alpha} \\
  &M_{\alpha\beta,\beta\alpha} + q =R \ddot{w}^1 + S \ddot{u}^1_{\alpha,\alpha} - T \ddot{w}^1_{,\alpha\alpha}
\end{empheq}
\end{subequations}
Setting $\varphi_{\alpha\beta}(z)=z\delta^K_{\alpha\beta}$ (where $\delta^K_{\alpha\beta}$ is the Kronecker symbol) in the present model gives the Mindlin--Reissner model. $P_{\alpha\beta}$ becomes symmetric and is equal to $M_{\alpha\beta}$, and $Q_{\alpha}$ are now the classical shear forces (previously denoted $Q^c_{\alpha}$ in this document). If $\gamma_{13,2}$ and $\gamma_{12,3}$ are summed, matrices $\mathbf{E}$, $\mathbf{F}$ and $\mathbf{G}$ can be reduced to $3\times3$ matrices. Then, $\mathbf{E}=\mathbf{A}$, $\mathbf{F}=\mathbf{B}$ and $\mathbf{G}=\mathbf{D}$ so that the $-w_{,\alpha\beta}$ and the $\gamma_{\alpha3,\beta}$ can be combined to give the classical symmetrical part $\psi_{\alpha\beta}=1/2(\phi_{\alpha,\beta}+\phi_{\beta,\alpha})$ of the gradient of rotations $\phi_{\alpha}=-w_{,\alpha}+\gamma_{\alpha3}$ of the Mindlin--Reissner model. The $10\times10$ stiffness matrix reduces to the classical $6\times6$ four-block membrane--bending behavior matrix. The shear matrix $\mathbf{H}$ is not null, and it can be enhanced by shear correction factors. The rotations second derivative with respect to time $\ddot{\phi}_{\alpha}=-\ddot{w}_{,\alpha}+\ddot{\gamma}_{\alpha3}$ will also appear. Note also that $U_{\alpha\beta}=S\delta^K_{\alpha\beta}$, $V_{\alpha\beta}=T\delta^K_{\alpha\beta}$ and $W_{\alpha\beta}=T\delta^K_{\alpha\beta}$. The system~\eqref{eq:behavior} can then be reduced, in a slightly different manner than in the previous case, with the help of~\eqref{eq:equilibrium1b}, leading to $5$ equations:
\begin{subequations}\label{eq:equilibriumMR}
\begin{empheq}[left=\empheqlbrace]{align}
  &N_{\alpha\beta,\beta}=R \ddot{u}^1_{\alpha} + S \ddot{\phi}_{\alpha} \\
  &M_{\alpha\beta,\beta}-Q_{\alpha}=S \ddot{u}^1_{\alpha} + T \ddot{\phi}_{\alpha} \\
	&Q_{\alpha,\alpha} + q = R \ddot{w}^1
\end{empheq}
\end{subequations}  
Reddy's original third-order theory~\cite{Reddy1984} is very efficient for a single layer plate because the kinematic assumption is very close to the three-dimensional elasticity solution. It can be simulated setting $\varphi_{\alpha\beta}(z)=(z-4/3z^3)\delta^K_{\alpha\beta}$ in the present model. This will work perfectly with the present formulation, but the link with Woodcock's work is not valid anymore, due to the presence of cubic terms in $z$. But this model is not adapted to high modulus ratios between adjacent layers because the warping functions are those of an homogeneous plate. 
\par
More sophisticated models with refined warping functions have been proposed. We shall pay attention only to those for which the number of unknowns does not depend on the number of layers. Among these models, we can cite the references~\cite{Pai1995,Kim2006}. These authors obtain warping functions for the general case, starting from undetermined cubic functions in each layer. Then they develop a (different) method to combine the layerwise polynomial functions into $4$ laminate warping functions. It is also obvious at this point that these warping functions can be put into our model. 
\subsection{Discussion based on numerical results}
This section intends to show the accuracy of the results for the present model and compare it to the other major models. 
\subsubsection{Numerical methods}

\paragraph{Navier's procedure}
This procedure is suitable for (specific) simply supported boundary conditions and for laminates with in-axis orthotropic plies. Each model is implemented in order to simulate the static and dynamic behavior of a square plate with boundary conditions $w=0$ on the four sides, $u_1=0$ on the $y=0$ and $y=a$ sides and $u_2=0$ on the $x=0$ and $x=a$ sides. With these assumptions, the generalized displacement field is:
\begin{equation}
	\label{eq:deplacementEE}
	\left\{
	\begin{array}{c}
		u_1 \\
		u_2 \\
		w \\
		\gamma_{13} \\
		\gamma_{23}
	\end{array}
	\right\}
	=
  \left\{
	\begin{array}{cll}
		u_1^{mn}         & \cos (m \pi x / a) & \sin (n \pi y / a)  \\
		u_2^{mn}         & \sin (m \pi x / a) & \cos (n \pi y / a)  \\
		w^{mn}           & \sin (m \pi x / a) & \sin (n \pi y / a)  \\
		\gamma_{13}^{mn} & \cos (m \pi x / a) & \sin (n \pi y / a)  \\
		\gamma_{23}^{mn} & \sin (m \pi x / a) & \cos (n \pi y / a)
	\end{array}
	\right\}
\end{equation}
where $m=1$ and $n=1$ for static analysis, or set to arbitrary $m$ and $n$ for the dynamic study of the corresponding mode.
There is no summation in these terms, which means that only one term is needed for each quantity. Then for a given $m$ and $n$, and for each model, the motion equations of section~\ref{sec:otherformulations} give a linear system for the coefficients $\{u_1^{mn},u_2^{mn}\dots\}$. For both cases the linear system is solved; for the static case, it directly gives the solution, whereas, for the dynamic case, a transfer function $w(\omega)$ is obtained. In this case, the first positive pole of this transfer function is the corresponding eigenfrequency of the $(m,n)$ mode.
\paragraph{Rayleigh--Ritz method}
In order to manage various boundary conditions and more general laminates, a numerical code based on the Rayleigh--Ritz method, previously described in references~\cite{Loredo2011,Castel2012}, has been updated to accept off-axis orthotropic plies. This code builds a stiffness matrix $\mymat{K}$ and a mass matrix $\mymat{M}$ from which we can compute eigenvalues, and then the natural angular frequencies. These angular frequencies are later nondimensionalized using formulas of the following type:
\begin{equation}
  \overline{\omega}=\omega \frac{a^2}{h} \sqrt{\frac{\rho}{E}}
  \label{eq:freqadim}
\end{equation}
where $h$ is the total thickness of the multilayer structure, $a$ is the side length, $\rho$ is a density and $E$ a Young's modulus. This generic formula must be adapted to present multi-material cases. The reader will find in the captions of tables~\ref{tab:results_hetero_reel},~\ref{tab:sand1},~\ref{tab:sand2},~\ref{tab:sand3} and~\ref{tab:sand4} the specific formulas which have been used for each case.
The basis used for all calculations with this code is the enriched trigonometric basis described by Beslin~\cite{Beslin1997633}. In the following, results obtained with the Rayleigh--Ritz method has been computed with a $30\times30$ basis. This order permits to ensure convergence up to $4$ significant digits.
\subsubsection{Models}
Different models are used for comparison in the following sections:
\begin{itemize}[--] \itemsep 4pt \topsep 0pt \parskip 0pt
  \item a 3D finite element model (3D FEM) which is taken as reference,
	\item the Love--Kirchhoff (L--K) plate model,
	\item the Mindlin--Reissner (M--R) plate model,
	\item the Mindlin--Reissner plate model with shear correction factors (SCF)\footnote{The SCF are computed with the method presented by Pai~\cite{Pai1995} and the Sun and Whitney displacement field.} (M--R with SCF), 
	\item the present model with warping functions of formula~\eqref{eq:warping_functions} obtained with the Sun and Whitney (S\:\&\:W) model,
	\item the present model with the warping functions obtained by Pai~\cite{Pai1995}.
\end{itemize}
\par
\subsubsection{Isotropic three layer plates with various Young's modulus ratios (Navier's method)}
\label{sec:ratiostudy}
In this section,  a simply supported three-layer plate made of two isotropic materials is studied. Deflection at the center of the plate and first natural frequency are given in tables~\ref{tab:results_hetero_reel} and~\ref{tab:results_hetero_reel2}. The outer layers $\ell=1$ and $\ell=3$ are made of the same material $M_s$ of Young's modulus $E_s$. The core layer, $\ell=2$ is made of a material $M_c$ of Young's modulus $E_c$. It is also assumed that $\rho_s=\rho_c$, $\nu_s=\nu_c=0.33$. The square plate has a side length~$a$, the length to thickness ratio is $a/h=400$, also, each layer has the same thickness $h_1=h_2=h_3$ and $h_1+h_2+h_3=h$. In table~\ref{tab:results_hetero_reel}, non dimensionless deflection is given for each model and for a varying ratio of $E_s/E_c$ from $1$ to $10^6$. In table~\ref{tab:results_hetero_reel2}, the first dimensionless natural frequency is given for each model and for the same varying parameters.

\begin{table}[htpb]
\centering
\small
\begin{tabular}{llllllll} \hline
  \multirow{2}{*}{Model}     &$E_s$/$E_c$&          &          &          &          &          &           \\ \cmidrule{2-8}
                             & $10^0$    & $10^1$   & $10^2$   & $10^3$   & $10^4$   & $10^5$   & $10^6$    \\ \hline
 L--K                        & 0.439105  & 0.454246 & 0.455818 & 0.455976 & 0.455992 & 0.455993 & 0.455993  \\
 M--R                        & 0.439118  & 0.454266 & 0.455838 & 0.455996 & 0.456012 & 0.456013 & 0.456014  \\
 M--R with SCF               & 0.439121  & 0.454299 & 0.456195 & 0.459587 & 0.491944 & 0.815363 & 4.049535  \\
 Present model with S\:\&\:W & 0.439118  & 0.454336 & 0.456683 & 0.464578 & 0.540778 & 1.200069 & 3.801913  \\
 Present model with Pai      & 0.439121  & 0.454312 & 0.456315 & 0.460775 & 0.503438 & 0.895972 & 3.008532  \\
 3D FEM                      & 0.439025  & 0.454293 & 0.456668 & 0.464573 & 0.540778 & 1.200050 & 3.801300  \\ \hline
 Relative difference$^*$     & 0.021 \%  & 0.010 \% & 0.003 \% & 0.001 \% & 0.000 \% & 0.002 \% & 0.016 \%   \\ \hline
\end{tabular}
\normalsize
\caption{Dimensionless deflection $\overline{w}=-10^{-5}(h E_s/a^2) w$ for the three layer $[M_s,M_c,M_s]$ isotropic laminate, with $a/h=400$ for different models and for various ratios $E_s/E_c$. $^*$Relative difference between the present model with the warping functions obtained with Sun and Whitney's model and the 3D FEM results.}

\label{tab:results_hetero_reel}
\end{table}
\begin{table}[htpb]
\centering
\small
\begin{tabular}{llllllll} \hline
  \multirow{2}{*}{Model}       &$E_s$/$E_c$&         &         &         &         &         &         \\ \cmidrule{2-8}
                               & $10^0$    & $10^1$  & $10^2$  & $10^3$  & $10^4$  & $10^5$  & $10^6$  \\ \hline
  L--K                         & 6.03637   & 5.93491 & 5.92467 & 5.92365 & 5.92354 & 5.92353 & 5.92353 \\
  M--R                         & 6.03628   & 5.93479 & 5.92454 & 5.92351 & 5.92341 & 5.92340 & 5.92340 \\
  M--R with SCF                & 6.03626   & 5.93457 & 5.92223 & 5.90033 & 5.70298 & 4.42980 & 1.98773 \\
  Present model with S\:\&\:W  & 6.03631   & 5.93435 & 5.91909 & 5.86858 & 5.43943 & 3.65141 & 2.05146 \\
  Present model with Pai       & 6.03629   & 5.93452 & 5.92147 & 5.89275 & 5.63755 & 4.22586 & 2.30609 \\
  3D FEM                       & 6.03673   & 5.93451 & 5.91909 & 5.86856 & 5.43942 & 3.65146 & 2.05160 \\ \hline
  Relative difference$^*$      &-0.007 \%  &-0.003 \%&-0.000 \%&-0.000 \%&-0.000 \%&-0.001 \%&-0.007 \%\\ \hline
\end{tabular}
\normalsize
\caption{Dimensionless fundamental natural frequency $\overline{\omega}=(\omega a^2/h)\sqrt{\rho/E_s}$ for the three layer $[M_s,M_c,M_s]$ isotropic laminate, with $a/h=400$ for different models and for various ratios $E_s/E_c$. $^*$Relative difference between the present model with the warping functions obtained with Sun and Whitney's model and the 3D FEM results.}
\label{tab:results_hetero_reel2}
\end{table}
\par
We can see from the results in tables~\ref{tab:results_hetero_reel} and~\ref{tab:results_hetero_reel2} that the Sun \& Whitney model gives the most accurate results. As expected, the classical models of Love-Kirchhoff and Mindlin--Reissner give the least accurate results, this can be explained since the Love-Kirchhoff model does not take into account the transverse shear and the Mindlin--Reissner model assumes the transverse shear strain to be constant over the thickness. It is interesting to note the difference between the results obtained with the Mindlin--Reissner plate model with shear correction factors obtained with Pai's method on Sun and Whitney's model and those obtained using directly warping functions obtained with Sun and Whitney's model. It is shown here that for multilayer structures exhibiting strong modulus ratios between layers, the present model is very accurate. This is the case for sandwich structures or constrained layer damping patches. Of course, for other cases, the warping functions are not adapted: for example, if this model is taken to study moderately thick plates without strong modulus ratios between layers, it would behave slightly better than the M--R model without SCF, but it would probably be less accurate than the M--R model with SCF. However, the shear correction factors only affect the shear behavior and does not affect mass terms, whereas the warping functions in present model modify both. This may explain that for dynamic behavior of damped plates at higher frequencies, the present model was found very accurate~\cite{Castel2012} and better than the M--R model with SCF. In addition, as told in the introduction, this model can be easily modified by changing the warping functions, hence, even if universal warping functions do not exist at this time, it is possible to adapt this model to many cases by choosing appropriate warping functions.
\subsubsection{Sandwich panels with anisotropic plies (Rayleigh--Ritz method)}
In this section two different sandwich square panels including off-axis orthotropic materials are studied. Dimensionless natural frequencies are compared to results from three dimensional finite element calculations. For both structures, length to thickness ratios varies from $a/h = 10$ to $a/h = 200$, and the core thickness to total thickness ratio is $h_c/h=0.88$. Face sheets are made of glass fiber and polyester resin and the core is made of HEREX-C70.130 foam; properties of both materials are defined in table~\ref{tab:matprop2}. These two sandwich examples have been found in reference~\cite{Nayak2002}, in which results in terms of natural frequencies have been given. However, comparison of these results with three-dimensional finite element simulation is difficult to do. Indeed, the authors have used specific boundary conditions, different for each sandwich, which are difficult to prescribe in a three dimensional model without perturbing the bending behavior due to the existing membrane--bending coupling for these two sandwiches (note that it was not the case for the previous study). Hence, in this study, classical simply supported boundary conditions have been prescribed: $w=0$ on all sides.
\par
Formula to nondimensionalize the frequencies is given in the caption of tables~\ref{tab:sand1},~\ref{tab:sand2},~\ref{tab:sand3}~and~\ref{tab:sand4} where $\rho_c$ and $E_c$ are respectively the density and Young's modulus of the core material. Finite element computations have been performed with a $50 \times 50 \times 21$ mesh of 20-node quadratic brick element.
\par
Tables~\ref{tab:sand1} and~\ref{tab:sand2} show that, in both cases, the model gives very satisfying results. Note that for the same precision, a $50\times50\times21$ mesh involves $672129$ degrees of freedom (DOF) whereas, for the same accuracy, with a base up to $30\times30$ the present code needs $4500$ DOF.
Since results for extremal values of $a/h = 10$ and $a/h = 200$ are in very good agreement with the finite element calculations, for intermediate values of $a/h$, only results of the present model are presented in tables~\ref{tab:sand3} and~\ref{tab:sand4}.
\begin{table}[htpb]
\centering
\small
\begin{tabular}{lccccccc}
            & $E_1$ (Pa)          & $E_2=E_3$ (Pa)      & $G_{12}=G_{13}$ (Pa)  & $G_{23}$ (Pa)  & $\nu_{12}=\nu_{13}$ & $\nu_{23}$ & $\rho$ (kg / m$^3$) \\ \hline
Face sheets &$24.51\times 10^9$   & $7.77\times 10^9$   & $3.34\times 10^9$ & $1.34\times 10^9$ & $0.246$              & $0.49$      & $1800$       \\
Core        &$103.63\times 10^6$  & $103.63\times 10^6$ & $50\times 10^6$       & $50\times 10^6$ & $0.32$             & $0.32$     & $130$       
\end{tabular}
\normalsize
\caption{Properties of the materials used in the sandwich panels}
\label{tab:matprop2}
\end{table}
\begin{table}[htpb]
\centering
\small
\begin{tabular}{cccccc}
Mode number &  Mode order  &     3D FEM     & Present model &      difference   \\
$a/h=10$    &    $(m,n)$   &  ($50\times50\times21$) & ($30\times30$) &  \%     \\ \hline
     1      &    $(1,1)$   &     14.073    &    14.057    &       -0.11       \\
     2      &    $(1,2)$   &     26.404    &    26.399    &       -0.02       \\ 
     3      &    $(2,1)$   &     27.008    &    27.009    &        0.00       \\ 
     4      &    $(2,2)$   &     35.184    &    35.219    &        0.10       \\ 
     5      &    $(1,3)$   &     40.177    &    40.272    &        0.24       \\ 
     6      &    $(3,1)$   &     40.954    &    41.063    &        0.27       \\ 
     7      &    $(2,3)$   &     46.418    &    46.586    &        0.36       \\ 
     8      &    $(3,2)$   &     46.772    &    46.943    &        0.37       \\ 
     9      &   membrane   &     51.249    &    51.413    &        0.32       \\ 
    10      &    $(1,4)$   &     53.990    &    54.328    &        0.63       \\ 
    11      &    $(4,1)$   &     54.969    &    55.320    &        0.64       \\ 
    12      &    $(3,3)$   &     55.652    &    56.000    &        0.63       \\ 
    13      &    $(2,4)$   &     58.759    &    59.202    &        0.75       \\ \hline  \\                   
$a/h=200$    &    $(m,n)$   &  ($50\times50\times21$) & ($30\times30$) &       \% \\ \hline
     1      &    $(1,1)$       &   22.236      &    22.250    &        0.06       \\
     2      &    $(1,2)$       &   54.913      &    54.933    &        0.04       \\
     3      &    $(2,1)$       &   61.955      &    61.973    &        0.03       \\ 
     4      &    $(2,2)$       &   88.494      &    88.543    &        0.05       \\ 
     5      &    $(1,3)$       &   111.82      &    111.84    &        0.02       \\ 
     6      &    $(3,1)$       &   129.62      &    129.63    &        0.01       \\ 
     7      &    $(2,3)$       &   140.45      &    140.52    &        0.05       \\ 
     8      &    $(3,2)$       &   152.27      &    152.33    &        0.04       \\
     9      &    $(1,4)$       &   191.30      &    191.30    &        0.00       \\
    10      &    $(3,3)$       &   197.47      &    197.58    &        0.06       \\
    11      &    $(2,4)$       &   217.01      &    217.07    &        0.03       \\
    12      &    $(4,1)$       &   223.37      &    223.37    &        0.00       \\
    13      &    $(4,2)$       &   243.94      &    243.99    &        0.02        
		
\end{tabular}
\normalsize
\caption{Dimensionless natural frequencies $\overline{\omega}=\omega (a^2/h) \sqrt{(\rho_c/E_c)}$ for a cross ply sequence of [0,90,0,core,0,90,0] square sandwich plate.}
\label{tab:sand1}
\end{table}
\begin{table}[htpb]
\centering
\small
\begin{tabular}{cccccc}
Mode number & Mode order &         3D FEM        & Present model  & difference   \\\hline
$a/h=10$    &  $(m,n)+$ shape & ($50\times50\times21$) & ($30\times30$) &  \% \\ \hline
     1      &  $(1,1)$        &     14.171           &     14.156    &     -0.10    \\
     2      &  $(1,2)$        &     26.624           &     26.621    &     -0.01    \\
     3      &  $(2,1)$        &     26.624           &     26.621    &     -0.01    \\
     4      &  $(2,2)$        &     35.418           &     35.462    &      0.13    \\
     5      &  $(2,2)$ diag.  &     40.251           &     40.341    &      0.22    \\
     6      &  $5=1+4$ star   &     40.437           &     40.536    &      0.24    \\
     7      &  $(2,3)$        &     46.784           &     46.966    &      0.39    \\
     8      &  $(3,2)$        &     46.784           &     46.966    &      0.39    \\
     9      &  $(4,1)$        &     54.161           &     54.487    &      0.60    \\
    10      &  $(1,4)$        &     54.161           &     54.487    &      0.60    \\
    11      &  $(3,3)$        &     55.990           &     56.360    &      0.66    \\
    12      &  membrane       &     56.859           &     57.120    &      0.46    \\             
    13      &  $8=0+8$ star   &     59.121           &     59.566    &      0.75    \\ \hline 
$a/h=200$   &  $(m,n)+$ shape &  ($50\times50\times21$) & ($30\times30$) &    \%   \\ 
\hline
     1      &    $(1,1)$      &     24.788    &    24.802    &        0.06       \\ 
     2      &    $(1,2)$      &     59.491    &    59.513    &        0.04       \\ 
     3      &    $(2,1)$      &     59.491    &    59.513    &        0.04       \\ 
     4      &    $(2,2)$      &     98.525    &    98.584    &        0.05       \\ 
     5      &    $(2,2)$ diag &     114.78    &    114.80    &        0.02       \\ 
     6      &    $5=1+4$ star &     114.81    &    114.83    &        0.02       \\ 
     7      &    $(2,3)$      &     156.92    &    156.99    &        0.05       \\ 
     8      &    $(3,2)$      &     156.92    &    156.99    &        0.05       \\ 
     9      &    $(1,4)$      &     190.88    &    190.89    &        0.00       \\ 
    10      &    $(4,1)$      &     190.88    &    190.89    &        0.00       \\ 
    11      &    $(3,3)$      &     219.42    &    219.55    &        0.06       \\ 
    12      &    $8=0+8$ star &     234.51    &    234.58    &        0.03       \\ 
    13      &    $(4,4)$ diag.&     234.62    &    234.69    &        0.03       \\ 
\end{tabular}
\normalsize
\caption{Dimensionless natural frequencies $\overline{\omega}=\omega (a^2/h) \sqrt{(\rho_c/E_c)}$ for an angle ply sequence of [45,-45,45,core,-45,45,-45] square sandwich plate.}
\label{tab:sand2}
\end{table}
\begin{table}[htpb]
\centering
\small
\begin{tabular}{cccc}
Mode number & $a/H = 20$ & $a/H = 50$ & $a/H = 100$ \\ \hline
1           &  18.515    &  21.274    &  21.971     \\
2           &  40.104    &  51.205    &  54.004     \\
3           &  42.653    &  56.976    &  60.765     \\
4           &  57.138    &  79.444    &  86.182     \\
5           &  67.968    &  99.217    &  108.79     \\
6           &  71.462    &  111.23    &  125.11     \\
7           &  80.020    &  121.24    &  135.51     \\
8           &  81.725    &  128.60    &  146.16     \\
9           &  97.118    &  159.02    &  183.12     \\
10          &  99.463    &  162.36    &  188.05     \\
11          &  100.81    &  176.30    &  206.40     \\
12          &  102.83    &  176.99    &  210.81     \\
13          &  106.29    &  190.55    &  229.28     \\
\end{tabular}
\normalsize
\caption{Dimensionless natural frequencies $\overline{\omega}=\omega (a^2/h) \sqrt{(\rho_c/E_c)}$ for a cross ply sequence of [0,90,0,core,0,90,0] square sandwich plate.}
\label{tab:sand3}
\end{table}
\begin{table}[htpb]
\centering
\small
\begin{tabular}{cccc}
Mode number & $a/H = 20$ & $a/H = 50$ & $a/H = 100$ \\ \hline
1           &  19.118    &  22.969    &  24.227     \\
2           &  41.030    &  54.037    &  57.983     \\
3           &  41.030    &  54.037    &  57.983     \\
4           &  58.485    &  84.972    &  94.644     \\
5           &  68.054    &  100.25    &  111.01     \\
6           &  68.478    &  100.58    &  111.13     \\
7           &  81.811    &  129.59    &  149.14     \\
8           &  81.811    &  129.59    &  149.14     \\
9           &  96.993    &  158.06    &  182.27     \\
10          &  96.993    &  158.06    &  182.27     \\
11          &  101.12    &  171.83    &  205.34     \\
12          &  107.47    &  184.02    &  219.95     \\
13          &  107.97    &  184.78    &  220.31     \\
\end{tabular}
\normalsize
\caption{Dimensionless natural frequencies $\overline{\omega}=\omega (a^2/h) \sqrt{(\rho_c/E_c)}$ for an angle ply sequence of [45,-45,45,core,-45,45,-45] square sandwich plate.}
\label{tab:sand4}
\end{table}

%% file: 04_Conclusion.tex
\section{Conclusion}
In this work, a multilayered plate model has been reformulated from Woodcock's work~\cite{Woodcock2008}. The pertinence of this model, especially for laminates with high modulus ratios between the layers, has been proven by comparison to three-dimensional finite element studies~\cite{Castel2012}. The reformulation has been motivated by several reasons. First, in the original work, there are missing terms in formulas for the energy coefficients, leading to an erroneous list of those coefficients. It is problematic because these energy coefficients are precisely the core of the method. Second, Woodcock's work is not presented with the standard notations of plate models, which makes it very difficult to understand and to situate relatively to other known models. For example, the displacement field, which is not explicit in the original work, has been made explicit in the present paper by means of warping functions.
\par
A mechanical behavior law involving a $10\times10$ stiffness matrix for the membrane / bending / shear warping behavior and their respective coupling, and a $2\times2$ stiffness matrix for the shear behavior, is given. This is suitable for static analysis. Complete equations of motion with generalized mass are also given. It has been proven that there are $55$ independent generalized stiffnesses components (or strain energy coefficients) and $14$ independent generalized mass (or kinetic energy coefficients).
\par
A complete link has been established between the original model and its new formulation, with all necessary proofs. In addition, the original work of Woodcock has been completed by additional formulas and detailed procedures to implement it in numerical simulations. It is ready to use in analytical works or in numerical simulations with Rayleigh--Ritz or finite element discretization for example.
\par
A discussion based on model formulations makes the place of the present model among classical multilayered plate models easy to see. In addition, a numerical study shows that this model (with the given warping functions) is very accurate when there is strong modulus ratios between the layers, which is the case for sandwich structures and also for plates damped with passive constrained layer patches.
\par
That permits to understand why the model is pertinent in the cases discussed above, and also to understand its limits, especially the one due to the hypothesis of constant transverse shear stresses, which is a poor hypothesis for the study of thick plates, except if shear correction coefficients are considered. This limit is not so problematic, because the present model can be, without effort, enhanced by a choice of  more refined  warping functions. For example, those issued from references~\cite{Pai1995,Kim2006} can be easily implemented for static or dynamic studies with the help of this work, which was definitely not the case with Woodcock's initial formulation.

%% file: 10_Appendix1.tex
\section{Calculation of Woodcock's layer coefficients\texorpdfstring{ $\alpha^{\ell}_{x},\alpha^{\ell}_{xy},\beta^{\ell}_{x}\dots$}{alpha, beta...}}\label{sec:LayerCoeffs}
In this section, the calculation  of the transfer matrix given in Woodcock's work at formula~(14) is performed. The notations used are those of section~\ref{sec:DisplacementField}.
\subsection{Link between the layer \texorpdfstring{$\ell+1$}{l+1} and the layer \texorpdfstring{$\ell$}{l}}
We start with the kinematic and ``static'' assumptions of formulas~\eqref{eq:cont_depl} and~\eqref{eq:cont_sigm}. Let us consider that $x$ and $y$ are fixed, hence they can be omitted in the terms. Enforcing condition~\eqref{eq:cont_depl} into~\eqref{eq:depl_W} leads to:  
\begin{gather}\label{eq:DeplCont}
  u_{\alpha}^{\ell+1}(z^{\ell+1})+(z^{\ell+1}-\zeta^\ell)(w^1_{,\alpha}-\gamma^{\ell+1}_{\alpha3})=u_{\alpha}^\ell(z^\ell)+(z^\ell-\zeta^\ell)(w^1_{,\alpha}-\gamma^\ell_{\alpha3})
\end{gather}
Let $\mathbf{C}_s^{\ell}$ be the transverse shear stiffness matrix of layer $\ell$ defined as:
\begin{gather}
	\mathbf{C}_s^{\ell}
	=
	\begin{bmatrix}
		   C^{\ell}_{1313}    &      C^{\ell}_{1323}    \\
		   C^{\ell}_{1323}    &      C^{\ell}_{2323}    
 	\end{bmatrix}
\end{gather}
Hence, with the ``static'' assumption~\eqref{eq:cont_sigm}:
\begin{gather}
  \begin{Bmatrix}
  	\gamma^{\ell+1}_{13} \\
	  \gamma^{\ell+1}_{23} \\
	\end{Bmatrix}
	=
	(\mathbf{C}_s^{\ell+1})^{-1}
  \begin{Bmatrix}
  	\sigma^{\ell+1}_{13} \\
	  \sigma^{\ell+1}_{23} \\
	\end{Bmatrix}
	=
	(\mathbf{C}_s^{\ell+1})^{-1}
  \begin{Bmatrix}
  	\sigma^{\ell}_{13} \\
	  \sigma^{\ell}_{23} \\
	\end{Bmatrix}
	=
	(\mathbf{C}_s^{\ell+1})^{-1}
	\mathbf{C}_s^{\ell}
  \begin{Bmatrix}
  	\gamma^{\ell}_{13} \\
	  \gamma^{\ell}_{23} \\
	\end{Bmatrix}
\end{gather}
Let us define:
\begin{gather}\label{eq:AlphaMatrix}
  \mathbf{\mathcal{A}}^{\ell+1}
  =
	(\mathbf{C}_s^{\ell+1})^{-1}
	\mathbf{C}_s^{\ell}
\end{gather}
hence:
\begin{gather}\label{eq:shearshear}
  \begin{Bmatrix}
  	\gamma^{\ell+1}_{13} \\
	  \gamma^{\ell+1}_{23} \\
	\end{Bmatrix}
	=
  \mathbf{\mathcal{A}}^{\ell+1}
  \begin{Bmatrix}
  	\gamma^{\ell}_{13} \\
	  \gamma^{\ell}_{23} \\
	\end{Bmatrix}
\end{gather}
For clarity we denote $\overline{u}^{\ell}_{\alpha}$ the mid-layer membrane displacements $u^\ell_{\alpha}(z^{\ell})$. Equation~\eqref{eq:DeplCont} can then be written:
\begin{gather}
  \begin{Bmatrix}
  	\overline{u}^{\ell+1}_1\\
	  \overline{u}^{\ell+1}_2 \\
	\end{Bmatrix}
	=
  \begin{Bmatrix}
  	\overline{u}^{\ell}_1 \\
	  \overline{u}^{\ell}_2 \\
	\end{Bmatrix}
	+
  (z^\ell-z^{\ell+1})
  \begin{Bmatrix}
  	w^1_{,1} \\
	  w^1_{,2} \\
	\end{Bmatrix}
  -
  (z^\ell-\zeta^{\ell})
  \begin{Bmatrix}
  	\gamma^{\ell}_{13} \\
	  \gamma^{\ell}_{23} \\
	\end{Bmatrix}
  +
  (z^{\ell+1}-\zeta^{\ell})
	\mathbf{\mathcal{A}}^{\ell+1}
  \begin{Bmatrix}
  	\gamma^{\ell}_{13} \\
	  \gamma^{\ell}_{23} \\
	\end{Bmatrix}
\end{gather}
And finally (minus signs are left to fit Woodcock's formulas):
\begin{gather}\label{eq:BetaGammaMatrix}
  \begin{Bmatrix}
  	\overline{u}^{\ell+1}_1\\
	  \overline{u}^{\ell+1}_2 \\
	\end{Bmatrix}
	=
  \:\mathbf{I}\:
	\begin{Bmatrix}
  	\overline{u}^{\ell}_1 \\
	  \overline{u}^{\ell}_2 \\
	\end{Bmatrix}
  -
  \mathbf{\mathcal{D}}^{\ell+1}
  \begin{Bmatrix}
  	-\gamma^{\ell}_{13} \\
	  -\gamma^{\ell}_{23} \\
	\end{Bmatrix}
	+
  \mathbf{\mathcal{B}}^{\ell+1}
  \begin{Bmatrix}
  	w^1_{,1} \\
	  w^1_{,2} \\
	\end{Bmatrix}
\end{gather}
with:
\begin{gather}\label{eq:BetaGammaMatrix2}
  \mathbf{\mathcal{B}}^{\ell+1}=(z^\ell-z^{\ell+1})\:\mathbf{I}\:
  \quad
	\text{and}
  \quad
  \mathbf{\mathcal{D}}^{\ell+1}=\frac{h^\ell}{2}\:\mathbf{I}\:+\frac{h^{\ell+1}}{2}\mathbf{\mathcal{A}}^{\ell+1}
\end{gather}
Equations~\eqref{eq:AlphaMatrix} and~\eqref{eq:BetaGammaMatrix} can be written~:
\begin{gather}\label{eq:recurrence1}
  \begin{Bmatrix}
	  \overline{u}_1^{\ell+1} \\
		\overline{u}_2^{\ell+1} \\
		-\gamma_{13}^{\ell+1} \\
		-\gamma_{23}^{\ell+1} \\
		w_{,1}^{\ell+1} \\
		w_{,2}^{\ell+1}
	\end{Bmatrix}
	=
	\begin{bmatrix}
		\:\mathbf{I}\: & -\mathbf{\mathcal{D}}^{\ell+1} & \mathbf{\mathcal{B}}^{\ell+1}  \\
		        0      & \mathbf{\mathcal{A}}^{\ell+1} &               0              \\
	 	        0      &               0               &        \:\mathbf{I}\:
 	\end{bmatrix}
	\begin{Bmatrix}
	  \overline{u}_1^{\ell} \\
		\overline{u}_2^{\ell} \\
		-\gamma_{13}^{\ell} \\
		-\gamma_{23}^{\ell} \\
		w_{,1}^{\ell} \\
		w_{,2}^{\ell}
	\end{Bmatrix}
\end{gather}
Let $\boldsymbol{\Upsilon}^{\ell}$ denote the vector involved in the right side of equation~\eqref{eq:recurrence1} and $\boldsymbol{\Omega}^{\ell+1}$ the corresponding block matrix: 
\begin{gather}\label{eq:recurrence2}
  \boldsymbol{\Upsilon}^{\ell+1} = \boldsymbol{\Omega}^{\ell+1} \boldsymbol{\Upsilon}^{\ell}
\end{gather}
\subsection{Link between the layer \texorpdfstring{$\ell$}{l} and the layer \texorpdfstring{$1$}{1} (for \texorpdfstring{$\ell>1$}{l>1})}
Let us try to compute a matrix which links the layer $\ell$ and the layer $1$:
\begin{gather}\label{eq:recurrence3}
  \boldsymbol{\Upsilon}^{\ell} = \boldsymbol{\Omega}^{\ell} \boldsymbol{\Omega}^{\ell-1} \dots \boldsymbol{\Omega}^2 \boldsymbol{\Upsilon}^{1}
	= \boldsymbol{\hat{\Omega}}^{\ell} \boldsymbol{\Upsilon}^{1}
\end{gather}
It is shown in the following lines that the matrix $\boldsymbol{\hat{\Omega}}^\ell$ can be written:  
\begin{gather}
	\boldsymbol{\hat{\Omega}}^\ell
	=
	\begin{bmatrix}
		\:\mathbf{I}\: & -\mathbf{\mathfrak{D}}^{\ell} & \mathbf{\mathfrak{B}}^{\ell}  \\
		        0      & \mathbf{\mathfrak{A}}^{\ell} &               0              \\
	 	        0      &               0             &        \:\mathbf{I}\:
 	\end{bmatrix}
\end{gather}
where:
\begin{subequations}\label{eq:recurrence4}
\begin{align}
	\mathbf{\mathfrak{A}}^{\ell} &= (\mathbf{C}_s^{\ell})^{-1} \mathbf{C}_s^{1} \\
	\mathbf{\mathfrak{D}}^{\ell} &= \frac{h^1}{2}\mathbf{\mathfrak{A}}^1+\sum_{m=2}^{\ell-1} h^m \mathbf{\mathfrak{A}}^m + \frac{h^{\ell}}{2}\mathbf{\mathfrak{A}}^{\ell}\\
	\mathbf{\mathfrak{B}}^{\ell} &= (z^1-z^{\ell}) \mathbf{\:I\:} = -z^{\ell} \mathbf{\:I\:}
\end{align}
\end{subequations}
Let us make the proof by recurrence. Setting $\ell=1$ into equations~\eqref{eq:AlphaMatrix} and~\eqref{eq:BetaGammaMatrix2} leads to:
\begin{subequations}
\begin{align}
 \mathbf{\mathfrak{A}}^{2} &= \mathbf{\mathcal{A}}^{2} = (\mathbf{C}_s^{2})^{-1} \mathbf{C}_s^{1} \\
 \mathbf{\mathfrak{D}}^{2} &= \mathbf{\mathcal{D}}^{2} = \frac{h^1}{2}\:\mathbf{I}\:+\frac{h^{2}}{2}\mathbf{\mathcal{A}}^{2} \\
 \mathbf{\mathfrak{B}}^{2} &= \mathbf{\mathcal{B}}^{2} = (z^1-z^{2})\:\mathbf{I}\: = -z^{2}\:\mathbf{I}\:
\end{align}
\end{subequations}
Writing $\boldsymbol{\hat{\Omega}}^{\ell+1}=\boldsymbol{\Omega}^{\ell+1}\boldsymbol{\hat{\Omega}}^{\ell}$ leads to:
\begin{align}		
	\boldsymbol{\hat{\Omega}}^{\ell+1}
	=
	\begin{bmatrix}
		\:\mathbf{I}\: & -\mathbf{\mathfrak{D}}^{\ell} - \mathbf{\mathcal{D}}^{\ell+1} \mathbf{\mathfrak{A}}^{\ell}      & \mathbf{\mathfrak{B}}^{\ell} + \mathbf{\mathcal{B}}^{\ell+1} \\
		        0      &                          \mathbf{\mathcal{A}}^{\ell+1} \mathbf{\mathfrak{A}}^{\ell}            &                                  0                           \\
	 	        0      &                                                     0                                          &                             \:\mathbf{I}\:
 	\end{bmatrix}
\end{align}
Calculation of the three non trivial blocks shows that the formulas~\eqref{eq:recurrence4} work also for the layer $\ell+1$,
\begin{subequations}
\begin{align}		
	\mathbf{\mathfrak{A}}^{\ell+1} &= \mathbf{\mathcal{A}}^{\ell+1} \mathbf{\mathfrak{A}}^{\ell} \nonumber \\
	                               &= (\mathbf{C}_s^{\ell+1})^{-1} \mathbf{C}_s^{\ell} (\mathbf{C}_s^{\ell})^{-1} \mathbf{C}_s^{1} \nonumber \\
																 &= (\mathbf{C}_s^{\ell+1})^{-1} \mathbf{C}_s^{1} \label{eq:blocA} \\
	\mathbf{\mathfrak{D}}^{\ell+1} &= \mathbf{\mathfrak{D}}^{\ell} + \mathbf{\mathcal{D}}^{\ell+1} \mathbf{\mathfrak{A}}^{\ell} \nonumber \\
	                               &= \frac{h^1}{2}\mathbf{\mathfrak{A}}^1+\sum_{m=2}^{\ell-1} h^m \mathbf{\mathfrak{A}}^m + \frac{h^{\ell}}{2}\mathbf{\mathfrak{A}}^{\ell} 
																  + \left(\frac{h^\ell}{2}\mathbf{\:I\:}+\frac{h^{\ell+1}}{2}\mathbf{\mathcal{A}}^{\ell+1}\right)\mathbf{\mathfrak{A}}^{\ell} \nonumber \\
															   &= \frac{h^1}{2}\mathbf{\mathfrak{A}}^1+\sum_{m=2}^{\ell} h^m \mathbf{\mathfrak{A}}^m + \frac{h^{\ell+1}}{2}\mathbf{\mathfrak{A}}^{\ell+1} \\
	\mathbf{\mathfrak{B}}^{\ell+1} &= \mathbf{\mathfrak{B}}^{\ell} + \mathbf{\mathcal{B}}^{\ell+1} \nonumber \\ 
	                               &= (z^1-z^{\ell}) \mathbf{\:I\:} + (z^{\ell}-z^{\ell+1}) \mathbf{\:I\:} \nonumber \\ 
	                               &= (z^1-z^{\ell+1}) \mathbf{\:I\:} = -z^{\ell+1} \mathbf{\:I\:} \label{eq:blocC}
\end{align}
\end{subequations}
that ends the proof. 
\par 
The $\boldsymbol{\hat{\Omega}}^\ell$ matrix contains terms which are here denoted in a slightly different manner than in Woodcock's work:
\begin{align}		\label{eq:recurrence}
	\boldsymbol{\hat{\Omega}}^\ell
	=
	\begin{bmatrix}
		\:\mathbf{I}\: & -\mathbf{\mathfrak{D}}^{\ell} & \mathbf{\mathfrak{B}}^{\ell}  \\
		        0      & \mathbf{\mathfrak{A}}^{\ell} &               0              \\
	 	        0      &               0             &        \:\mathbf{I}\:
 	\end{bmatrix}
	=
	\begin{bmatrix}
		   1    &      0     & \delta_{xx}^{\ell} & \delta_{xy}^{\ell} & \beta^{\ell} &        0       \\
		   0    &      1     & \delta_{yx}^{\ell} & \delta_{yy}^{\ell} &        0       & \beta^{\ell} \\
		   0    &      0     & \alpha_{xx}^{\ell} & \alpha_{xy}^{\ell} &        0       &        0       \\
		   0    &      0     & \alpha_{yx}^{\ell} & \alpha_{yy}^{\ell} &        0       &        0       \\
		   0    &      0     &          0         &          0         &        1       &        0       \\
	 	   0    &      0     &          0         &	         0         &        0       &        1 
 	\end{bmatrix}
\end{align}
This notation permits to write generic formulas like~\eqref{eq:deltaWoodcock} for example. 
The correspondence with Woodcock's notation and the matrix in formula~(14) of Woodcock's article~\cite{Woodcock2008} can be obtained following the rules:
\begin{itemize}[--] \topsep 0pt \itemsep 4pt \parskip 0pt
	\item replace superscripts $\ell$ by $n-1$,
	\item replace $\delta_{xx}^{n-1}$ by $\gamma_{x}^{n-1}$ and $\delta_{yy}^{n-1}$ by $\gamma_{y}^{n-1}$,
	\item replace $\alpha_{xx}^{n-1}$ by $\alpha_{x}^{n-1}$ and $\alpha_{yy}^{n-1}$ by $\alpha_{y}^{n-1}$,
	\item put a $\beta_x^{n-1}$ and a $\beta_y^{n-1}$ instead of the two (identical) $\beta^{n-1}$,
	\item swap rows and columns of matrix following $(1,2,3,4,5,6)\rightarrow(5,3,1,6,4,2)$
\end{itemize}

%% file: 11_Appendix2.tex
\section{Corrected lists of Woodcock's energies coefficients}
\label{sec:EnergyCoeffs}
\subsection{Important note}
Below are listed the corrected $69$ Woodcock's strain energy coefficients and the corrected $18$ kinetic energy coefficients as they were defined in the article~\cite{Woodcock2008}. 
\par
All the calculation has been done with the help of a symbolic computation software. The coefficients have been computed following Woodcock's procedure: once established that the transfer matrix of equation~\eqref{eq:recurrence} could link each layer to the first one, strain and kinetic energies are expressed in function of the first layer displacement field. Integration of these strain and kinetic energies over the thickness of the plate leads to functionals given in reference~\cite{Woodcock2008} at formulas~(24) and~(18), and in Appendix~B. Woodcock found that these functional has respectively $69$ and $13$ coefficients, some of them being equal. Reproducing Woodcock's procedure, the computation shows that mistakes occur in the given lists. A quick examination of these lists shows that none of the coefficients exhibit a $\delta_{xy}$ or a $\alpha_{xy}$ term which are present in the transfer matrix of formula (14) of reference~\cite{Woodcock2008}.
\par
Another proof is given in this work, mainly in \ref{sec:Proof}, in which these same coefficients are calculated with a different method involving the warping functions of formula~\eqref{eq:warping_functions}.
\subsection{Preliminary remarks}
Some preliminary remarks must be done at this time:
\begin{enumerate}
	\item Energy coefficients have been given in Woodcock's work using the two indices Voigt notation for the fourth-order stiffness tensor, so we choose to keep here this index convention. We shall recall here that the change is made according to the well known rule $(11,22,33,23,13,12)\leftrightarrow(1,2,3,4,5,6)$. 
	\item Note that in this document a $\ell$ superscript has been chosen for the layer index instead of the $n$ superscript in Woodcock's formulas. Note also that a particular choice of numbering in Woodcock's work has lead to a mix of $n$ and $n-1$ superscript which hides the true nature of inner terms of coefficients, which are simply ``weights'' for each layer into generalized stiffnesses or mass. 
	\item The density $\mu^n$ of the $n^\text{th}$ layer in Woodcock's work has been changed with $\rho^\ell$ in the following formulas. 
\end{enumerate}
\par
Some equalities between coefficients are listed in equations~\eqref{eq:equallambdas} and~\eqref{eq:equaldeltas}. It can be verified that these relations obtained by symmetry considerations on the energies are confirmed in the lists below. Note that there remain only $55$ and $14$ independent coefficients for respectively strain and kinetic energies, in the most general case.
\subsection{Definition of coefficients \texorpdfstring{$\lambda_i$}{lambda}}
\begin{flalign}
  \input{Lambdas.txt}
\end{flalign}
\subsection{Definition of coefficients \texorpdfstring{$\delta_i$}{delta}}
\begin{flalign}
  \input{Deltas.txt}
\end{flalign}

%% file: Lambdas.txt
\nonumber\lambda_1 =\sum_{\ell=1}^n\bigg[& \left( \frac{1}{12}  \left( {h}^{\ell} \right) ^3 +{h}^{\ell} \left( {\beta}^{\ell} \right) ^2  \right) Q_{11}^{\ell}\bigg]\phantom{\frac{1}{2}} \\ \medskip
\nonumber\lambda_2 =\sum_{\ell=1}^n\bigg[& \left( \frac{1}{12}  \left( \alpha_{x}^{\ell} \right) ^2  \left( {h}^{\ell} \right) ^3 + \left( \gamma_{x}^{\ell} \right) ^2 {h}^{\ell} \right) Q_{11}^{\ell}+ \left( \frac{1}{6} \alpha_{x}^{\ell}\alpha_{yx}^{\ell} \left( {h}^{\ell} \right) ^3 +2 \gamma_{x}^{\ell}\delta_{yx}^{\ell}{h}^{\ell} \right) Q_{16}^{\ell} \\ \nonumber + & \left( \frac{1}{12}  \left( \alpha_{yx}^{\ell} \right) ^2  \left( {h}^{\ell} \right) ^3 + \left( \delta_{yx}^{\ell} \right) ^2 {h}^{\ell} \right) Q_{66}^{\ell}\bigg]\phantom{\frac{1}{2}} \\ \medskip
\nonumber\lambda_3 =\sum_{\ell=1}^n\bigg[&{h}^{\ell}Q_{11}^{\ell}\bigg]\phantom{\frac{1}{2}} \\ \medskip
\nonumber\lambda_4 =\sum_{\ell=1}^n\bigg[& \left( \frac{1}{6} \alpha_{x}^{\ell} \left( {h}^{\ell} \right) ^3 +2 \gamma_{x}^{\ell}{h}^{\ell}{\beta}^{\ell} \right) Q_{11}^{\ell}+ \left( \frac{1}{6} \alpha_{yx}^{\ell} \left( {h}^{\ell} \right) ^3 +2 \delta_{yx}^{\ell}{h}^{\ell}{\beta}^{\ell} \right) Q_{16}^{\ell}\bigg]\phantom{\frac{1}{2}} \\ \medskip
\nonumber\lambda_5 =\sum_{\ell=1}^n\bigg[&2 {h}^{\ell}{\beta}^{\ell}Q_{11}^{\ell}\bigg]\phantom{\frac{1}{2}} \\ \medskip
\nonumber\lambda_6 =\sum_{\ell=1}^n\bigg[&2 \gamma_{x}^{\ell}{h}^{\ell}Q_{11}^{\ell}+2 \delta_{yx}^{\ell}{h}^{\ell}Q_{16}^{\ell}\bigg]\phantom{\frac{1}{2}} \\ \medskip
\nonumber\lambda_7 =\sum_{\ell=1}^n\bigg[& \left( \frac{1}{12}  \left( {h}^{\ell} \right) ^3 +{h}^{\ell} \left( {\beta}^{\ell} \right) ^2  \right) Q_{22}^{\ell}\bigg]\phantom{\frac{1}{2}} \\ \medskip
\nonumber\lambda_8 =\sum_{\ell=1}^n\bigg[& \left( \frac{1}{12}  \left( \alpha_{y}^{\ell} \right) ^2  \left( {h}^{\ell} \right) ^3 + \left( \gamma_{y}^{\ell} \right) ^2 {h}^{\ell} \right) Q_{22}^{\ell}+ \left( \frac{1}{6} \alpha_{y}^{\ell}\alpha_{xy}^{\ell} \left( {h}^{\ell} \right) ^3 +2 \gamma_{y}^{\ell}\delta_{xy}^{\ell}{h}^{\ell} \right) Q_{26}^{\ell} \\ \nonumber + &\left( \frac{1}{12}  \left( \alpha_{xy}^{\ell} \right) ^2  \left( {h}^{\ell} \right) ^3 + \left( \delta_{xy}^{\ell} \right) ^2 {h}^{\ell} \right) Q_{66}^{\ell}\bigg]\phantom{\frac{1}{2}} \\ \medskip
\nonumber\lambda_9 =\sum_{\ell=1}^n\bigg[&{h}^{\ell}Q_{22}^{\ell}\bigg]\phantom{\frac{1}{2}} \\ \medskip
\nonumber\lambda_{10}=\sum_{\ell=1}^n\bigg[& \left( \frac{1}{6} \alpha_{y}^{\ell} \left( {h}^{\ell} \right) ^3 +2 \gamma_{y}^{\ell}{h}^{\ell}{\beta}^{\ell} \right) Q_{22}^{\ell}+ \left( \frac{1}{6} \alpha_{xy}^{\ell} \left( {h}^{\ell} \right) ^3 +2 \delta_{xy}^{\ell}{h}^{\ell}{\beta}^{\ell} \right) Q_{26}^{\ell}\bigg]\phantom{\frac{1}{2}} \\ \medskip
\nonumber\lambda_{11}=\sum_{\ell=1}^n\bigg[&2 {h}^{\ell}{\beta}^{\ell}Q_{22}^{\ell}\bigg]\phantom{\frac{1}{2}} \\ \medskip
\nonumber\lambda_{12}=\sum_{\ell=1}^n\bigg[&2 \gamma_{y}^{\ell}{h}^{\ell}Q_{22}^{\ell}+2 \delta_{xy}^{\ell}{h}^{\ell}Q_{26}^{\ell}\bigg]\phantom{\frac{1}{2}} \\ \medskip
\nonumber\lambda_{13}=\sum_{\ell=1}^n\bigg[& \left( \frac{1}{6}  \left( {h}^{\ell} \right) ^3 +2 {h}^{\ell} \left( {\beta}^{\ell} \right) ^2  \right) Q_{12}^{\ell}\bigg]\phantom{\frac{1}{2}} \\ \medskip
\nonumber\lambda_{14}=\sum_{\ell=1}^n\bigg[& \left( \frac{1}{6} \alpha_{y}^{\ell} \left( {h}^{\ell} \right) ^3 +2 \gamma_{y}^{\ell}{h}^{\ell}{\beta}^{\ell} \right) Q_{12}^{\ell}+ \left( \frac{1}{6} \alpha_{xy}^{\ell} \left( {h}^{\ell} \right) ^3 +2 \delta_{xy}^{\ell}{h}^{\ell}{\beta}^{\ell} \right) Q_{16}^{\ell}\bigg]\phantom{\frac{1}{2}} \\ \medskip
\nonumber\lambda_{15}=\sum_{\ell=1}^n\bigg[&2 {h}^{\ell}{\beta}^{\ell}Q_{12}^{\ell}\bigg]\phantom{\frac{1}{2}} \\ \medskip
\nonumber\lambda_{16}=\sum_{\ell=1}^n\bigg[& \left( \frac{1}{6} \alpha_{x}^{\ell} \left( {h}^{\ell} \right) ^3 +2 \gamma_{x}^{\ell}{h}^{\ell}{\beta}^{\ell} \right) Q_{12}^{\ell}+ \left( \frac{1}{6} \alpha_{yx}^{\ell} \left( {h}^{\ell} \right) ^3 +2 \delta_{yx}^{\ell}{h}^{\ell}{\beta}^{\ell} \right) Q_{26}^{\ell}\bigg]\phantom{\frac{1}{2}} \\ \medskip
\nonumber\lambda_{17}=\sum_{\ell=1}^n\bigg[& \left( \frac{1}{6} \alpha_{x}^{\ell}\alpha_{y}^{\ell} \left( {h}^{\ell} \right) ^3 +2 \gamma_{x}^{\ell}\gamma_{y}^{\ell}{h}^{\ell} \right) Q_{12}^{\ell}+ \left( \frac{1}{6} \alpha_{x}^{\ell}\alpha_{xy}^{\ell} \left( {h}^{\ell} \right) ^3 +2 \gamma_{x}^{\ell}\delta_{xy}^{\ell}{h}^{\ell} \right) Q_{16}^{\ell} \\ \nonumber + &\left( \frac{1}{6} \alpha_{y}^{\ell}\alpha_{yx}^{\ell} \left( {h}^{\ell} \right) ^3 +2 \gamma_{y}^{\ell}\delta_{yx}^{\ell}{h}^{\ell} \right) Q_{26}^{\ell}+ \left( \frac{1}{6} \alpha_{xy}^{\ell}\alpha_{yx}^{\ell} \left( {h}^{\ell} \right) ^3 +2 \delta_{xy}^{\ell}\delta_{yx}^{\ell}{h}^{\ell} \right) Q_{66}^{\ell}\bigg]\phantom{\frac{1}{2}} \\ \medskip
\nonumber\lambda_{18}=\sum_{\ell=1}^n\bigg[&2 \gamma_{x}^{\ell}{h}^{\ell}Q_{12}^{\ell}+2 \delta_{yx}^{\ell}{h}^{\ell}Q_{26}^{\ell}\bigg]\phantom{\frac{1}{2}} \\ \medskip
\nonumber\lambda_{19}=\sum_{\ell=1}^n\bigg[&2 {h}^{\ell}{\beta}^{\ell}Q_{12}^{\ell}\bigg]\phantom{\frac{1}{2}} \\ \medskip
\nonumber\lambda_{20}=\sum_{\ell=1}^n\bigg[&2 \gamma_{y}^{\ell}{h}^{\ell}Q_{12}^{\ell}+2 \delta_{xy}^{\ell}{h}^{\ell}Q_{16}^{\ell}\bigg]\phantom{\frac{1}{2}} \\ \medskip
\nonumber\lambda_{21}=\sum_{\ell=1}^n\bigg[&2 {h}^{\ell}Q_{12}^{\ell}\bigg]\phantom{\frac{1}{2}} \\ \medskip
\nonumber\lambda_{22}=\sum_{\ell=1}^n\bigg[& \left( \frac{1}{3}  \left( {h}^{\ell} \right) ^3 +4 {h}^{\ell} \left( {\beta}^{\ell} \right) ^2  \right) Q_{66}^{\ell}\bigg]\phantom{\frac{1}{2}} \\ \medskip
\nonumber\lambda_{23}=\sum_{\ell=1}^n\bigg[& \left( \frac{1}{12}  \left( \alpha_{yx}^{\ell} \right) ^2  \left( {h}^{\ell} \right) ^3 + \left( \delta_{yx}^{\ell} \right) ^2 {h}^{\ell} \right) Q_{22}^{\ell}+ \left( \frac{1}{6} \alpha_{x}^{\ell}\alpha_{yx}^{\ell} \left( {h}^{\ell} \right) ^3 +2 \gamma_{x}^{\ell}\delta_{yx}^{\ell}{h}^{\ell} \right) Q_{26}^{\ell} \\ \nonumber + &\left( \frac{1}{12}  \left( \alpha_{x}^{\ell} \right) ^2  \left( {h}^{\ell} \right) ^3 + \left( \gamma_{x}^{\ell} \right) ^2 {h}^{\ell} \right) Q_{66}^{\ell}\bigg]\phantom{\frac{1}{2}} \\ \medskip
\nonumber\lambda_{24}=\sum_{\ell=1}^n\bigg[& \left( \frac{1}{12}  \left( \alpha_{xy}^{\ell} \right) ^2  \left( {h}^{\ell} \right) ^3 + \left( \delta_{xy}^{\ell} \right) ^2 {h}^{\ell} \right) Q_{11}^{\ell}+ \left( \frac{1}{6} \alpha_{y}^{\ell}\alpha_{xy}^{\ell} \left( {h}^{\ell} \right) ^3 +2 \gamma_{y}^{\ell}\delta_{xy}^{\ell}{h}^{\ell} \right) Q_{16}^{\ell} \\ \nonumber + &\left( \frac{1}{12}  \left( \alpha_{y}^{\ell} \right) ^2  \left( {h}^{\ell} \right) ^3 + \left( \gamma_{y}^{\ell} \right) ^2 {h}^{\ell} \right) Q_{66}^{\ell}\bigg]\phantom{\frac{1}{2}} \\ \medskip
\nonumber\lambda_{25}=\sum_{\ell=1}^n\bigg[&{h}^{\ell}Q_{66}^{\ell}\bigg]\phantom{\frac{1}{2}} \\ \medskip
\nonumber\lambda_{26}=\sum_{\ell=1}^n\bigg[&{h}^{\ell}Q_{66}^{\ell}\bigg]\phantom{\frac{1}{2}} \\ \medskip
\nonumber\lambda_{27}=\sum_{\ell=1}^n\bigg[& \left( \frac{1}{3} \alpha_{yx}^{\ell} \left( {h}^{\ell} \right) ^3 +4 \delta_{yx}^{\ell}{h}^{\ell}{\beta}^{\ell} \right) Q_{26}^{\ell}+ \left( \frac{1}{3} \alpha_{x}^{\ell} \left( {h}^{\ell} \right) ^3 +4 \gamma_{x}^{\ell}{h}^{\ell}{\beta}^{\ell} \right) Q_{66}^{\ell}\bigg]\phantom{\frac{1}{2}} \\ \medskip
\nonumber\lambda_{28}=\sum_{\ell=1}^n\bigg[& \left( \frac{1}{3} \alpha_{xy}^{\ell} \left( {h}^{\ell} \right) ^3 +4 \delta_{xy}^{\ell}{h}^{\ell}{\beta}^{\ell} \right) Q_{16}^{\ell}+ \left( \frac{1}{3} \alpha_{y}^{\ell} \left( {h}^{\ell} \right) ^3 +4 \gamma_{y}^{\ell}{h}^{\ell}{\beta}^{\ell} \right) Q_{66}^{\ell}\bigg]\phantom{\frac{1}{2}} \\ \medskip
\nonumber\lambda_{29}=\sum_{\ell=1}^n\bigg[&4 {h}^{\ell}{\beta}^{\ell}Q_{66}^{\ell}\bigg]\phantom{\frac{1}{2}} \\ \medskip
\nonumber\lambda_{30}=\sum_{\ell=1}^n\bigg[&4 {h}^{\ell}{\beta}^{\ell}Q_{66}^{\ell}\bigg]\phantom{\frac{1}{2}} \\ \medskip
\nonumber\lambda_{31}=\sum_{\ell=1}^n\bigg[& \left( \frac{1}{6} \alpha_{xy}^{\ell}\alpha_{yx}^{\ell} \left( {h}^{\ell} \right) ^3 +2 \delta_{xy}^{\ell}\delta_{yx}^{\ell}{h}^{\ell} \right) Q_{12}^{\ell}+ \left( \frac{1}{6} \alpha_{x}^{\ell}\alpha_{xy}^{\ell} \left( {h}^{\ell} \right) ^3 +2 \gamma_{x}^{\ell}\delta_{xy}^{\ell}{h}^{\ell} \right) Q_{16}^{\ell} \\ \nonumber + &\left( \frac{1}{6} \alpha_{y}^{\ell}\alpha_{yx}^{\ell} \left( {h}^{\ell} \right) ^3 +2 \gamma_{y}^{\ell}\delta_{yx}^{\ell}{h}^{\ell} \right) Q_{26}^{\ell}+ \left( \frac{1}{6} \alpha_{x}^{\ell}\alpha_{y}^{\ell} \left( {h}^{\ell} \right) ^3 +2 \gamma_{x}^{\ell}\gamma_{y}^{\ell}{h}^{\ell} \right) Q_{66}^{\ell}\bigg]\phantom{\frac{1}{2}} \\ \medskip
\nonumber\lambda_{32}=\sum_{\ell=1}^n\bigg[&2 \delta_{yx}^{\ell}{h}^{\ell}Q_{26}^{\ell}+2 \gamma_{x}^{\ell}{h}^{\ell}Q_{66}^{\ell}\bigg]\phantom{\frac{1}{2}} \\ \medskip
\nonumber\lambda_{33}=\sum_{\ell=1}^n\bigg[&2 \delta_{yx}^{\ell}{h}^{\ell}Q_{26}^{\ell}+2 \gamma_{x}^{\ell}{h}^{\ell}Q_{66}^{\ell}\bigg]\phantom{\frac{1}{2}} \\ \medskip
\nonumber\lambda_{34}=\sum_{\ell=1}^n\bigg[&2 \delta_{xy}^{\ell}{h}^{\ell}Q_{16}^{\ell}+2 \gamma_{y}^{\ell}{h}^{\ell}Q_{66}^{\ell}\bigg]\phantom{\frac{1}{2}} \\ \medskip
\nonumber\lambda_{35}=\sum_{\ell=1}^n\bigg[&2 \delta_{xy}^{\ell}{h}^{\ell}Q_{16}^{\ell}+2 \gamma_{y}^{\ell}{h}^{\ell}Q_{66}^{\ell}\bigg]\phantom{\frac{1}{2}} \\ \medskip
\nonumber\lambda_{36}=\sum_{\ell=1}^n\bigg[&2 {h}^{\ell}Q_{66}^{\ell}\bigg]\phantom{\frac{1}{2}} \\ \medskip
\nonumber\lambda_{37}=\sum_{\ell=1}^n\bigg[& \left( \alpha_{yx}^{\ell} \right) ^2 {h}^{\ell}C_{44}^{\ell}+2 \alpha_{x}^{\ell}\alpha_{yx}^{\ell}{h}^{\ell}C_{45}^{\ell}+ \left( \alpha_{x}^{\ell} \right) ^2 {h}^{\ell}C_{55}^{\ell}\bigg]\phantom{\frac{1}{2}} \\ \medskip
\nonumber\lambda_{38}=\sum_{\ell=1}^n\bigg[& \left( \alpha_{y}^{\ell} \right) ^2 {h}^{\ell}C_{44}^{\ell}+2 \alpha_{y}^{\ell}\alpha_{xy}^{\ell}{h}^{\ell}C_{45}^{\ell}+ \left( \alpha_{xy}^{\ell} \right) ^2 {h}^{\ell}C_{55}^{\ell}\bigg]\phantom{\frac{1}{2}} \\ \medskip
\nonumber\lambda_{39}=\sum_{\ell=1}^n\bigg[& \left( \frac{1}{3}  \left( {h}^{\ell} \right) ^3 +4 {h}^{\ell} \left( {\beta}^{\ell} \right) ^2  \right) Q_{16}^{\ell}\bigg]\phantom{\frac{1}{2}} \\ \medskip
\nonumber\lambda_{40}=\sum_{\ell=1}^n\bigg[& \left( \frac{1}{6} \alpha_{yx}^{\ell} \left( {h}^{\ell} \right) ^3 +2 \delta_{yx}^{\ell}{h}^{\ell}{\beta}^{\ell} \right) Q_{12}^{\ell}+ \left( \frac{1}{6} \alpha_{x}^{\ell} \left( {h}^{\ell} \right) ^3 +2 \gamma_{x}^{\ell}{h}^{\ell}{\beta}^{\ell} \right) Q_{16}^{\ell}\bigg]\phantom{\frac{1}{2}} \\ \medskip
\nonumber\lambda_{41}=\sum_{\ell=1}^n\bigg[&2 {h}^{\ell}{\beta}^{\ell}Q_{16}^{\ell}\bigg]\phantom{\frac{1}{2}} \\ \medskip
\nonumber\lambda_{42}=\sum_{\ell=1}^n\bigg[& \left( \frac{1}{3} \alpha_{x}^{\ell} \left( {h}^{\ell} \right) ^3 +4 \gamma_{x}^{\ell}{h}^{\ell}{\beta}^{\ell} \right) Q_{16}^{\ell}+ \left( \frac{1}{3} \alpha_{yx}^{\ell} \left( {h}^{\ell} \right) ^3 +4 \delta_{yx}^{\ell}{h}^{\ell}{\beta}^{\ell} \right) Q_{66}^{\ell}\bigg]\phantom{\frac{1}{2}} \\ \medskip
\nonumber\lambda_{43}=\sum_{\ell=1}^n\bigg[& \left( \frac{1}{6} \alpha_{x}^{\ell}\alpha_{yx}^{\ell} \left( {h}^{\ell} \right) ^3 +2 \gamma_{x}^{\ell}\delta_{yx}^{\ell}{h}^{\ell} \right) Q_{12}^{\ell}+ \left( \frac{1}{6}  \left( \alpha_{x}^{\ell} \right) ^2  \left( {h}^{\ell} \right) ^3 +2  \left( \gamma_{x}^{\ell} \right) ^2 {h}^{\ell} \right) Q_{16}^{\ell} \\ \nonumber + &\left( \frac{1}{6}  \left( \alpha_{yx}^{\ell} \right) ^2  \left( {h}^{\ell} \right) ^3 +2  \left( \delta_{yx}^{\ell} \right) ^2 {h}^{\ell} \right) Q_{26}^{\ell}+ \left( \frac{1}{6} \alpha_{x}^{\ell}\alpha_{yx}^{\ell} \left( {h}^{\ell} \right) ^3 +2 \gamma_{x}^{\ell}\delta_{yx}^{\ell}{h}^{\ell} \right) Q_{66}^{\ell}\bigg]\phantom{\frac{1}{2}} \\ \medskip
\nonumber\lambda_{44}=\sum_{\ell=1}^n\bigg[&2 \gamma_{x}^{\ell}{h}^{\ell}Q_{16}^{\ell}+2 \delta_{yx}^{\ell}{h}^{\ell}Q_{66}^{\ell}\bigg]\phantom{\frac{1}{2}} \\ \medskip
\nonumber\lambda_{45}=\sum_{\ell=1}^n\bigg[&4 {h}^{\ell}{\beta}^{\ell}Q_{16}^{\ell}\bigg]\phantom{\frac{1}{2}} \\ \medskip
\nonumber\lambda_{46}=\sum_{\ell=1}^n\bigg[&2 \delta_{yx}^{\ell}{h}^{\ell}Q_{12}^{\ell}+2 \gamma_{x}^{\ell}{h}^{\ell}Q_{16}^{\ell}\bigg]\phantom{\frac{1}{2}} \\ \medskip
\nonumber\lambda_{47}=\sum_{\ell=1}^n\bigg[&2 {h}^{\ell}Q_{16}^{\ell}\bigg]\phantom{\frac{1}{2}} \\ \medskip
\nonumber\lambda_{48}=\sum_{\ell=1}^n\bigg[& \left( \frac{1}{6} \alpha_{xy}^{\ell} \left( {h}^{\ell} \right) ^3 +2 \delta_{xy}^{\ell}{h}^{\ell}{\beta}^{\ell} \right) Q_{11}^{\ell}+ \left( \frac{1}{6} \alpha_{y}^{\ell} \left( {h}^{\ell} \right) ^3 +2 \gamma_{y}^{\ell}{h}^{\ell}{\beta}^{\ell} \right) Q_{16}^{\ell}\bigg]\phantom{\frac{1}{2}} \\ \medskip
\nonumber\lambda_{49}=\sum_{\ell=1}^n\bigg[&2 {h}^{\ell}{\beta}^{\ell}Q_{16}^{\ell}\bigg]\phantom{\frac{1}{2}} \\ \medskip
\nonumber\lambda_{50}=\sum_{\ell=1}^n\bigg[& \left( \frac{1}{6} \alpha_{x}^{\ell}\alpha_{xy}^{\ell} \left( {h}^{\ell} \right) ^3 +2 \gamma_{x}^{\ell}\delta_{xy}^{\ell}{h}^{\ell} \right) Q_{11}^{\ell}+ \bigg( \frac{1}{6} \alpha_{x}^{\ell}\alpha_{y}^{\ell} \left( {h}^{\ell} \right) ^3 +\frac{1}{6} \alpha_{xy}^{\ell}\alpha_{yx}^{\ell} \left( {h}^{\ell} \right) ^3  \\ \nonumber + &2 \gamma_{x}^{\ell}\gamma_{y}^{\ell}{h}^{\ell}+2 \delta_{xy}^{\ell}\delta_{yx}^{\ell}{h}^{\ell} \bigg) Q_{16}^{\ell}+ \left( \frac{1}{6} \alpha_{y}^{\ell}\alpha_{yx}^{\ell} \left( {h}^{\ell} \right) ^3 +2 \gamma_{y}^{\ell}\delta_{yx}^{\ell}{h}^{\ell} \right) Q_{66}^{\ell}\bigg]\phantom{\frac{1}{2}} \\ \medskip
\nonumber\lambda_{51}=\sum_{\ell=1}^n\bigg[&2 \gamma_{x}^{\ell}{h}^{\ell}Q_{16}^{\ell}+2 \delta_{yx}^{\ell}{h}^{\ell}Q_{66}^{\ell}\bigg]\phantom{\frac{1}{2}} \\ \medskip
\nonumber\lambda_{52}=\sum_{\ell=1}^n\bigg[&2 \delta_{xy}^{\ell}{h}^{\ell}Q_{11}^{\ell}+2 \gamma_{y}^{\ell}{h}^{\ell}Q_{16}^{\ell}\bigg]\phantom{\frac{1}{2}} \\ \medskip
\nonumber\lambda_{53}=\sum_{\ell=1}^n\bigg[&2 {h}^{\ell}Q_{16}^{\ell}\bigg]\phantom{\frac{1}{2}} \\ \medskip
\nonumber\lambda_{54}=\sum_{\ell=1}^n\bigg[& \left( \frac{1}{3}  \left( {h}^{\ell} \right) ^3 +4 {h}^{\ell} \left( {\beta}^{\ell} \right) ^2  \right) Q_{26}^{\ell}\bigg]\phantom{\frac{1}{2}} \\ \medskip
\nonumber\lambda_{55}=\sum_{\ell=1}^n\bigg[& \left( \frac{1}{6} \alpha_{yx}^{\ell} \left( {h}^{\ell} \right) ^3 +2 \delta_{yx}^{\ell}{h}^{\ell}{\beta}^{\ell} \right) Q_{22}^{\ell}+ \left( \frac{1}{6} \alpha_{x}^{\ell} \left( {h}^{\ell} \right) ^3 +2 \gamma_{x}^{\ell}{h}^{\ell}{\beta}^{\ell} \right) Q_{26}^{\ell}\bigg]\phantom{\frac{1}{2}} \\ \medskip
\nonumber\lambda_{56}=\sum_{\ell=1}^n\bigg[&2 {h}^{\ell}{\beta}^{\ell}Q_{26}^{\ell}\bigg]\phantom{\frac{1}{2}} \\ \medskip
\nonumber\lambda_{57}=\sum_{\ell=1}^n\bigg[& \left( \frac{1}{3} \alpha_{y}^{\ell} \left( {h}^{\ell} \right) ^3 +4 \gamma_{y}^{\ell}{h}^{\ell}{\beta}^{\ell} \right) Q_{26}^{\ell}+ \left( \frac{1}{3} \alpha_{xy}^{\ell} \left( {h}^{\ell} \right) ^3 +4 \delta_{xy}^{\ell}{h}^{\ell}{\beta}^{\ell} \right) Q_{66}^{\ell}\bigg]\phantom{\frac{1}{2}} \\ \medskip
\nonumber\lambda_{58}=\sum_{\ell=1}^n\bigg[& \left( \frac{1}{6} \alpha_{y}^{\ell}\alpha_{yx}^{\ell} \left( {h}^{\ell} \right) ^3 +2 \gamma_{y}^{\ell}\delta_{yx}^{\ell}{h}^{\ell} \right) Q_{22}^{\ell}+ \bigg( \frac{1}{6} \alpha_{x}^{\ell}\alpha_{y}^{\ell} \left( {h}^{\ell} \right) ^3 +\frac{1}{6} \alpha_{xy}^{\ell}\alpha_{yx}^{\ell} \left( {h}^{\ell} \right) ^3  \\ \nonumber + &2 \gamma_{x}^{\ell}\gamma_{y}^{\ell}{h}^{\ell}+2 \delta_{xy}^{\ell}\delta_{yx}^{\ell}{h}^{\ell} \bigg) Q_{26}^{\ell}+ \left( \frac{1}{6} \alpha_{x}^{\ell}\alpha_{xy}^{\ell} \left( {h}^{\ell} \right) ^3 +2 \gamma_{x}^{\ell}\delta_{xy}^{\ell}{h}^{\ell} \right) Q_{66}^{\ell}\bigg]\phantom{\frac{1}{2}} \\ \medskip
\nonumber\lambda_{59}=\sum_{\ell=1}^n\bigg[&2 \gamma_{y}^{\ell}{h}^{\ell}Q_{26}^{\ell}+2 \delta_{xy}^{\ell}{h}^{\ell}Q_{66}^{\ell}\bigg]\phantom{\frac{1}{2}} \\ \medskip
\nonumber\lambda_{60}=\sum_{\ell=1}^n\bigg[&4 {h}^{\ell}{\beta}^{\ell}Q_{26}^{\ell}\bigg]\phantom{\frac{1}{2}} \\ \medskip
\nonumber\lambda_{61}=\sum_{\ell=1}^n\bigg[&2 \delta_{yx}^{\ell}{h}^{\ell}Q_{22}^{\ell}+2 \gamma_{x}^{\ell}{h}^{\ell}Q_{26}^{\ell}\bigg]\phantom{\frac{1}{2}} \\ \medskip
\nonumber\lambda_{62}=\sum_{\ell=1}^n\bigg[&2 {h}^{\ell}Q_{26}^{\ell}\bigg]\phantom{\frac{1}{2}} \\ \medskip
\nonumber\lambda_{63}=\sum_{\ell=1}^n\bigg[& \left( \frac{1}{6} \alpha_{xy}^{\ell} \left( {h}^{\ell} \right) ^3 +2 \delta_{xy}^{\ell}{h}^{\ell}{\beta}^{\ell} \right) Q_{12}^{\ell}+ \left( \frac{1}{6} \alpha_{y}^{\ell} \left( {h}^{\ell} \right) ^3 +2 \gamma_{y}^{\ell}{h}^{\ell}{\beta}^{\ell} \right) Q_{26}^{\ell}\bigg]\phantom{\frac{1}{2}} \\ \medskip
\nonumber\lambda_{64}=\sum_{\ell=1}^n\bigg[&2 {h}^{\ell}{\beta}^{\ell}Q_{26}^{\ell}\bigg]\phantom{\frac{1}{2}} \\ \medskip
\nonumber\lambda_{65}=\sum_{\ell=1}^n\bigg[& \left( \frac{1}{6} \alpha_{y}^{\ell}\alpha_{xy}^{\ell} \left( {h}^{\ell} \right) ^3 +2 \gamma_{y}^{\ell}\delta_{xy}^{\ell}{h}^{\ell} \right) Q_{12}^{\ell}+ \left( \frac{1}{6}  \left( \alpha_{xy}^{\ell} \right) ^2  \left( {h}^{\ell} \right) ^3 +2  \left( \delta_{xy}^{\ell} \right) ^2 {h}^{\ell} \right) Q_{16}^{\ell} \\ \nonumber + & \left( \frac{1}{6}  \left( \alpha_{y}^{\ell} \right) ^2  \left( {h}^{\ell} \right) ^3 +2  \left( \gamma_{y}^{\ell} \right) ^2 {h}^{\ell} \right) Q_{26}^{\ell}+ \left( \frac{1}{6} \alpha_{y}^{\ell}\alpha_{xy}^{\ell} \left( {h}^{\ell} \right) ^3 +2 \gamma_{y}^{\ell}\delta_{xy}^{\ell}{h}^{\ell} \right) Q_{66}^{\ell}\bigg]\phantom{\frac{1}{2}} \\ \medskip
\nonumber\lambda_{66}=\sum_{\ell=1}^n\bigg[&2 \gamma_{y}^{\ell}{h}^{\ell}Q_{26}^{\ell}+2 \delta_{xy}^{\ell}{h}^{\ell}Q_{66}^{\ell}\bigg]\phantom{\frac{1}{2}} \\ \medskip
\nonumber\lambda_{67}=\sum_{\ell=1}^n\bigg[&2 \delta_{xy}^{\ell}{h}^{\ell}Q_{12}^{\ell}+2 \gamma_{y}^{\ell}{h}^{\ell}Q_{26}^{\ell}\bigg]\phantom{\frac{1}{2}} \\ \medskip
\nonumber\lambda_{68}=\sum_{\ell=1}^n\bigg[&2 {h}^{\ell}Q_{26}^{\ell}\bigg]\phantom{\frac{1}{2}} \\ \medskip
\nonumber\lambda_{69}=\sum_{\ell=1}^n\bigg[&2 \alpha_{y}^{\ell}\alpha_{yx}^{\ell}{h}^{\ell}C_{44}^{\ell}+ \left( 2 \alpha_{x}^{\ell}\alpha_{y}^{\ell}{h}^{\ell}+2 \alpha_{xy}^{\ell}\alpha_{yx}^{\ell}{h}^{\ell} \right) C_{45}^{\ell}+2 \alpha_{x}^{\ell}\alpha_{xy}^{\ell}{h}^{\ell}C_{55}^{\ell}\bigg]\phantom{\frac{1}{2}} \\ \medskip

%% file: Deltas.txt
\nonumber\delta_1 =\sum_{\ell=1}^n\bigg[& \left( \frac{1}{12}  \left( {h}^{\ell} \right) ^3 +{h}^{\ell} \left( {\beta}^{\ell} \right) ^2  \right) \rho^{\ell}\bigg]\phantom{\frac{1}{2}} \\ \medskip
\nonumber\delta_2 =\sum_{\ell=1}^n\bigg[& \left( \frac{1}{12}  \left( \alpha_{x}^{\ell} \right) ^2  \left( {h}^{\ell} \right) ^3 +\frac{1}{12}  \left( \alpha_{yx}^{\ell} \right) ^2  \left( {h}^{\ell} \right) ^3 + \left( \gamma_{x}^{\ell} \right) ^2 {h}^{\ell}+ \left( \delta_{yx}^{\ell} \right) ^2 {h}^{\ell} \right) \rho^{\ell}\bigg]\phantom{\frac{1}{2}} \\ \medskip
\nonumber\delta_3 =\sum_{\ell=1}^n\bigg[&{h}^{\ell}\rho^{\ell}\bigg]\phantom{\frac{1}{2}} \\ \medskip
\nonumber\delta_4 =\sum_{\ell=1}^n\bigg[& \left( \frac{1}{6} \alpha_{x}^{\ell} \left( {h}^{\ell} \right) ^3 +2 \gamma_{x}^{\ell}{h}^{\ell}{\beta}^{\ell} \right) \rho^{\ell}\bigg]\phantom{\frac{1}{2}} \\ \medskip
\nonumber\delta_5 =\sum_{\ell=1}^n\bigg[&2 {h}^{\ell}{\beta}^{\ell}\rho^{\ell}\bigg]\phantom{\frac{1}{2}} \\ \medskip
\nonumber\delta_6 =\sum_{\ell=1}^n\bigg[&2 \gamma_{x}^{\ell}{h}^{\ell}\rho^{\ell}\bigg]\phantom{\frac{1}{2}} \\ \medskip
\nonumber\delta_7 =\sum_{\ell=1}^n\bigg[& \left( \frac{1}{12}  \left( {h}^{\ell} \right) ^3 +{h}^{\ell} \left( {\beta}^{\ell} \right) ^2  \right) \rho^{\ell}\bigg]\phantom{\frac{1}{2}} \\ \medskip
\nonumber\delta_8 =\sum_{\ell=1}^n\bigg[& \left( \frac{1}{12}  \left( \alpha_{y}^{\ell} \right) ^2  \left( {h}^{\ell} \right) ^3 +\frac{1}{12}  \left( \alpha_{xy}^{\ell} \right) ^2  \left( {h}^{\ell} \right) ^3 + \left( \gamma_{y}^{\ell} \right) ^2 {h}^{\ell}+ \left( \delta_{xy}^{\ell} \right) ^2 {h}^{\ell} \right) \rho^{\ell}\bigg]\phantom{\frac{1}{2}} \\ \medskip
\nonumber\delta_9 =\sum_{\ell=1}^n\bigg[&{h}^{\ell}\rho^{\ell}\bigg]\phantom{\frac{1}{2}} \\ \medskip
\nonumber\delta_{10}=\sum_{\ell=1}^n\bigg[& \left( \frac{1}{6} \alpha_{y}^{\ell} \left( {h}^{\ell} \right) ^3 +2 \gamma_{y}^{\ell}{h}^{\ell}{\beta}^{\ell} \right) \rho^{\ell}\bigg]\phantom{\frac{1}{2}} \\ \medskip
\nonumber\delta_{11}=\sum_{\ell=1}^n\bigg[&2 {h}^{\ell}{\beta}^{\ell}\rho^{\ell}\bigg]\phantom{\frac{1}{2}} \\ \medskip
\nonumber\delta_{12}=\sum_{\ell=1}^n\bigg[&2 \gamma_{y}^{\ell}{h}^{\ell}\rho^{\ell}\bigg]\phantom{\frac{1}{2}} \\ \medskip
\nonumber\delta_{13}=\sum_{\ell=1}^n\bigg[&{h}^{\ell}\rho^{\ell}\bigg]\phantom{\frac{1}{2}} \\ \medskip
\nonumber\delta_{14}=\sum_{\ell=1}^n\bigg[& \left( \frac{1}{6} \alpha_{x}^{\ell}\alpha_{xy}^{\ell} \left( {h}^{\ell} \right) ^3 +\frac{1}{6} \alpha_{y}^{\ell}\alpha_{yx}^{\ell} \left( {h}^{\ell} \right) ^3 +2 \gamma_{x}^{\ell}\delta_{xy}^{\ell}{h}^{\ell}+2 \gamma_{y}^{\ell}\delta_{yx}^{\ell}{h}^{\ell} \right) \rho^{\ell}\bigg]\phantom{\frac{1}{2}} \\ \medskip
\nonumber\delta_{15}=\sum_{\ell=1}^n\bigg[& \left( \frac{1}{6} \alpha_{xy}^{\ell} \left( {h}^{\ell} \right) ^3 +2 \delta_{xy}^{\ell}{h}^{\ell}{\beta}^{\ell} \right) \rho^{\ell}\bigg]\phantom{\frac{1}{2}} \\ \medskip
\nonumber\delta_{16}=\sum_{\ell=1}^n\bigg[& \left( \frac{1}{6} \alpha_{yx}^{\ell} \left( {h}^{\ell} \right) ^3 +2 \delta_{yx}^{\ell}{h}^{\ell}{\beta}^{\ell} \right) \rho^{\ell}\bigg]\phantom{\frac{1}{2}} \\ \medskip
\nonumber\delta_{17}=\sum_{\ell=1}^n\bigg[&2 \delta_{yx}^{\ell}{h}^{\ell}\rho^{\ell}\bigg]\phantom{\frac{1}{2}} \\ \medskip
\nonumber\delta_{18}=\sum_{\ell=1}^n\bigg[&2 \delta_{xy}^{\ell}{h}^{\ell}\rho^{\ell}\bigg]\phantom{\frac{1}{2}} \\ \medskip

%% file: 12_Appendix3.tex
\section{Proof of the link between stiffness and mass matrices and Woodcock's coefficients}
\label{sec:Proof}
\subsection{Preliminary results}
We recall here that the height of each layer is written $h^{\ell}=\zeta^{\ell}-\zeta^{\ell-1}$ and that with the help of formulas~\eqref{eq:blocC} and~\eqref{eq:recurrence} it can be seen that $(\zeta^{\ell}+\zeta^{\ell-1})/2=z^{\ell}=-\beta^{\ell}$ is the layer mid-plane elevation. Let us first calculate:
\begin{subequations}
\begin{align}
\label{eq:intdz}
\int_{\zeta^{\ell-1}}^{\zeta^{\ell}} \text{d}z 
 &= h^{\ell} \\ 
\label{eq:intzdz}
\int_{\zeta^{\ell-1}}^{\zeta^{\ell}} z \text{d}z 
 &= \frac{1}{2}((\zeta^{\ell})^2-(\zeta^{\ell-1})^2) = h^{\ell}z^{\ell} = h^{\ell}(-\beta^{\ell})  \\
\nonumber
\label{eq:intz2dz}
\int_{\zeta^{\ell-1}}^{\zeta^{\ell}} z^2 \text{d}z 
 &= \frac{1}{3}((\zeta^{\ell})^3-(\zeta^{\ell-1})^3) = \frac{1}{12}((\zeta^{\ell})^3-(\zeta^{\ell-1})^3)+\frac{1}{4}((\zeta^{\ell})^3-(\zeta^{\ell-1})^3) \\
\nonumber
 &= \frac{1}{12}(\zeta^{\ell}-\zeta^{\ell-1})^3+\frac{1}{4}((\zeta^{\ell})^3-(\zeta^{\ell-1})^3+(\zeta^{\ell})^2\zeta^{\ell-1}-\zeta^{\ell}(\zeta^{\ell-1})^2) \\
 &= \frac{1}{12}(h^{\ell})^3+h^{\ell}(z^{\ell})^2 = \frac{1}{12}(h^{\ell})^3+h^{\ell}(\beta^{\ell})^2 
\end{align}
\end{subequations}
These above three formulas help to calculate:
\begin{subequations}
\begin{align}
	\label{eq:intzmoinsbetadz}
	\int_{\zeta^{\ell-1}}^{\zeta^{\ell}} (z-z^\ell)\text{d}z 
	 &= h^{\ell}(z^{\ell}-z^\ell) = 0 \\
	\label{eq:intzmoinsbetazdz}
	\int_{\zeta^{\ell-1}}^{\zeta^{\ell}} (z-z^\ell)z\text{d}z 
	 &=\frac{1}{12}(h^{\ell})^3+h^{\ell}(z^{\ell})^2-h^{\ell}(z^{\ell})^2=\frac{1}{12}(h^{\ell})^3\\
	\label{eq:intzmoinsbeta2dz}
	\int_{\zeta^{\ell-1}}^{\zeta^{\ell}} (z-z^\ell)^2\text{d}z 
	 &=\frac{1}{12}(h^{\ell})^3+h^{\ell}(z^{\ell})^2-2h^{\ell}(z^{\ell})^2+(z^{\ell})^2h^{\ell}=\frac{1}{12}(h^{\ell})^3
\end{align}
\end{subequations}
Another important preliminary result consists in the transformation of the following integral on a sum over the layers, written in a particular manner which leads to easier calculations on the following steps:
\begin{align}
  \nonumber
  \int_0^z S_{\alpha3\gamma3}(\zeta)\text{d}\zeta
	&=
	S^1_{\alpha3\gamma3}(\zeta^1-0)+\sum_{m=2}^{\ell-1} S^m_{\alpha3\gamma3}(\zeta^m-\zeta^{m-1})+S^\ell_{\alpha3\gamma3}(z-\zeta^{\ell-1}) \\ \label{eq:intS} 
	&=
	S^1_{\alpha3\gamma3}\frac{h^1}{2}+\sum_{m=2}^{\ell-1} S^m_{\alpha3\gamma3}h^m+S^\ell_{\alpha3\gamma3}\frac{h^\ell}{2}+S^\ell_{\alpha3\gamma3}(z-z^\ell)
\end{align}
\subsection{Integrals of the warping functions}
In this section, for clarity, the notations of equation~\eqref{eq:recurrence} for Woodcock's coefficients are used. The link with Woodcock's notations is explained just after equation~\eqref{eq:recurrence}.
\paragraph{Calculation of the $E_{\alpha\beta\gamma\delta}$} It requires the calculation of:
\begin{gather}
	\int_{\zeta^{\ell-1}}^{\zeta^{\ell}}\varphi_{\alpha\beta}(z)\text{d}z
	=
	4C_{\gamma3\beta3}(z^1)
	\int_{\zeta^{\ell-1}}^{\zeta^{\ell}}
	\left(
		\int_0^z S_{\alpha3\gamma3}(\zeta)\text{d}\zeta
	\right)
	\text{d}z
\end{gather}
according to preliminary results of formulas~\eqref{eq:intdz},~\eqref{eq:intzmoinsbetadz} and~\eqref{eq:intS}:
\begin{align}
	\int_{\zeta^{\ell-1}}^{\zeta^{\ell}}\varphi_{\alpha\beta}(z)\text{d}z
	=
	4C_{\gamma3\beta3}(z^1)
	\left(
		S^1_{\alpha3\gamma3}\frac{h^1}{2}+\sum_{m=2}^{\ell-1} S^m_{\alpha3\gamma3}h^m+S^\ell_{\alpha3\gamma3}\frac{h^{\ell}}{2}
	\right)
  h^{\ell}
\end{align}
\par
One can see now with the help of \ref{sec:LayerCoeffs} that the integrals of the warping functions over the layer $\ell$ are linked with Woodcock's coefficients $\delta^{\ell}_{\alpha\beta}$:
\begin{align}\label{eq:deltaWoodcock}
	\int_{\zeta^{\ell-1}}^{\zeta^{\ell}}\varphi_{\alpha\beta}(z)\text{d}z
	=
  -\delta^{\ell}_{\alpha\beta} h^{\ell}
\end{align}
\paragraph{Calculation of the $F_{\alpha\beta\gamma\delta}$} It requires the calculation of:
\begin{gather}
	\int_{\zeta^{\ell-1}}^{\zeta^{\ell}}\varphi_{\alpha\beta}(z)z\text{d}z
	=
	4C_{\gamma3\beta3}(z^1)
	\int_{\zeta^{\ell-1}}^{\zeta^{\ell}}
	\left(
		\int_0^z S_{\alpha3\gamma3}(\zeta)\text{d}\zeta
	\right)
	z\text{d}z
\end{gather}
according to preliminary results of formulas~\eqref{eq:intzdz},~\eqref{eq:intzmoinsbetazdz} and~\eqref{eq:intS}:
\begin{align}
	\int_{\zeta^{\ell-1}}^{\zeta^{\ell}}\varphi_{\alpha\beta}(z)z\text{d}z
	=
	4C_{\gamma3\beta3}(z^1)
	\left[
	\left(
		S^1_{\alpha3\gamma3}\frac{h^1}{2}+\sum_{m=2}^{\ell-1} S^m_{\alpha3\gamma3}h^m + S^\ell_{\alpha3\gamma3}\frac{h^{\ell}}{2}
	\right)
  (-h^{\ell}\beta^{\ell})
	+
		S^\ell_{\alpha3\gamma3}\frac{(h^{\ell})^3}{12}
	\right]
\end{align}
\par
One can see now with the help of \ref{sec:LayerCoeffs} that the first moments of the warping functions over the layer $\ell$ are linked with Woodcock's coefficients $\delta^{\ell}_{\alpha\beta}$, $\beta^{\ell}$ and $\alpha^{\ell}_{\alpha\beta}$:
\begin{align}\label{eq:deltabetaWoodcock}
	\int_{\zeta^{\ell-1}}^{\zeta^{\ell}}\varphi_{\alpha\beta}(z)z\text{d}z
	=
  \delta^{\ell}_{\alpha\beta} \beta^{\ell} h^{\ell} + \alpha_{\alpha\beta}^{\ell} \frac{(h^{\ell})^3}{12}
\end{align}
\paragraph{Calculation of the $G_{\alpha\beta\gamma\delta}$} It requires the calculation of:
\begin{gather}
	\int_{\zeta^{\ell-1}}^{\zeta^{\ell}}\varphi_{\alpha\beta}(z)\varphi_{\gamma\delta}(z)\text{d}z
	=
	16C_{\lambda3\beta3}(z^1)C_{\mu3\delta3}(z^1)
	\int_{\zeta^{\ell-1}}^{\zeta^{\ell}}
	\left(
		\int_0^z S_{\alpha3\lambda3}(\zeta)\text{d}\zeta
		\int_0^z S_{\gamma3\mu3}(\zeta)\text{d}\zeta
	\right)
	\text{d}z
\end{gather}
according to preliminary results of formulas~\eqref{eq:intdz},~\eqref{eq:intzmoinsbetadz},~\eqref{eq:intzmoinsbeta2dz} and~\eqref{eq:intS}:
\begin{align}
  \nonumber
	\int_{\zeta^{\ell-1}}^{\zeta^{\ell}}\varphi_{\alpha\beta}(z)\varphi_{\gamma\delta}(z)\text{d}z
	=
	&16C_{\lambda3\beta3}(z^1)C_{\mu3\delta3}(z^1)
	\int_{\zeta^{\ell-1}}^{\zeta^{\ell}}
	\Bigg[ \\ \nonumber
		&\left(
		  S^1_{\alpha3\lambda3}\frac{h^1}{2}+\sum_{m=2}^{\ell-1} S^m_{\alpha3\lambda3}h^m+S^\ell_{\alpha3\lambda3}\frac{h^\ell}{2}
	  \right)
    \left(
		  S^1_{\gamma3\mu3}\frac{h^1}{2}+\sum_{m=2}^{\ell-1} S^m_{\gamma3\mu3}h^m+S^\ell_{\gamma3\mu3}\frac{h^\ell}{2}
	  \right) \\   \nonumber
		+
		&\left(
		  S^1_{\alpha3\lambda3}\frac{h^1}{2}+\sum_{m=2}^{\ell-1} S^m_{\alpha3\lambda3}h^m+S^\ell_{\alpha3\lambda3}\frac{h^\ell}{2}
	  \right) S^\ell_{\gamma3\mu3} (z-z^\ell) \\ \nonumber
		+
    &\left(
		  S^1_{\gamma3\mu3}\frac{h^1}{2}+\sum_{m=2}^{\ell-1} S^m_{\gamma3\mu3}h^m+S^\ell_{\gamma3\mu3}\frac{h^\ell}{2}
	  \right) S^\ell_{\alpha3\lambda3}(z-z^\ell)\\ 
		+
		&S^\ell_{\gamma3\mu3} S^\ell_{\alpha3\lambda3}(z-z^\ell)^2
	\Bigg]
	\text{d}z
\end{align}
One can see now with the help of \ref{sec:LayerCoeffs} that this integral over the layer $\ell$ is linked with Woodcock's coefficients $\delta^{\ell}_{\alpha\beta}$ and $\alpha^{\ell}_{\alpha\beta}$:
\begin{gather}\label{eq:delta2ampha2Woodcock}
	\int_{\zeta^{\ell-1}}^{\zeta^{\ell}}\varphi_{\alpha\beta}(z)\varphi_{\gamma\delta}(z)\text{d}z
	=
	\delta_{\alpha\beta}\delta_{\gamma\delta}h^\ell + \alpha_{\alpha\beta}\alpha_{\gamma\delta}\frac{(h^\ell)^3}{12}
\end{gather}
\subsection{Calculation of matrices \texorpdfstring{$\mymat{A},\mymat{B},\mymat{D},\mymat{E},\mymat{F},\mymat{G},\mymat{H}$}{A, B, D, E, F, G, H} and \texorpdfstring{$\mymat{\boldsymbol{\Xi}}$}{Xi}}
\paragraph{Matrices $\mymat{A}$, $\mymat{B}$ and $\mymat{D}$}
The calculation of terms of these matrices is classical. Formula~\eqref{eq:generalized stiffnesses1} shows that results can be written directly in a matrix form:
\begin{gather}
  \{\mymat{A},\mymat{B},\mymat{D}\}=\int_{\zeta^0}^{\zeta^{n}}\mymat{Q}\{1,z,z^2\}\text{d}z=\sum_{\ell=1}^n\mymat{Q}^\ell\{h^{\ell},-h^{\ell}\beta^{\ell},\frac{(h^{\ell})^3}{12}+h^{\ell}(\beta^{\ell})^2\}
\end{gather}
One can see that these formulas corresponds to Woodcock's coefficients $\lambda_i$ with $i\in (3,9,21,25,47,62)$ for $\mymat{A}$, to $-\lambda_i$ with $i\in (5,11,15,29,45,60)$ for $\mymat{B}$ and to $\lambda_i$ with $i\in (1,7,13,22,39,54)$ for $\mymat{D}$. Note that some Woodcock's coefficients must be divided by $2$ or by $4$ according to the correspondence shown in formula~\eqref{eq:behaviorWoodcock}.  
\paragraph{Matrix $\mymat{E}$}
Formulas~\eqref{eq:generalized stiffnesses1} and~\eqref{eq:deltaWoodcock} give:
\begin{gather}
  E_{\alpha\beta\mu\delta}=\int^{\zeta^n}_{\zeta^0} Q_{\alpha\beta\gamma\delta}\varphi_{\gamma\mu}(z)\text{d}z=-\sum_{\ell=1}^n Q^{\ell}_{\alpha\beta\gamma\delta}\delta^{\ell}_{\gamma\mu}h^{\ell}
\end{gather}
This formula gives Woodcock's coefficients $-\lambda_i$ with $i\in (6,12,18,20,32,34,44,46,52,59,61,67)$. Let us finish the demonstration for some of them, changing the notations on the fly to fit Woodcock's coefficients:
\begin{subequations}
\begin{align}
  E_{1111}&=-\sum_{\ell=1}^n (Q^{\ell}_{1111}\delta^{\ell}_{11}+Q^{\ell}_{1121}\delta^{\ell}_{21})h^{\ell}=-\sum_{\ell=1}^n (Q^{\ell}_{11}\gamma^{\ell}_{x}+Q^{\ell}_{16}\delta^{\ell}_{yx})h^{\ell}=-\frac{\lambda_6}{2} \\
  E_{2211}&=-\sum_{\ell=1}^n (Q^{\ell}_{2211}\delta^{\ell}_{11}+Q^{\ell}_{2221}\delta^{\ell}_{21})h^{\ell}=-\sum_{\ell=1}^n (Q^{\ell}_{12}\gamma^{\ell}_{x}+Q^{\ell}_{26}\delta^{\ell}_{yx})h^{\ell}=-\frac{\lambda_{18}}{2} \\
  E_{1122}&=-\sum_{\ell=1}^n (Q^{\ell}_{1112}\delta^{\ell}_{12}+Q^{\ell}_{1122}\delta^{\ell}_{22})h^{\ell}=-\sum_{\ell=1}^n (Q^{\ell}_{16}\delta^{\ell}_{xy}+Q^{\ell}_{12}\gamma^{\ell}_{y})h^{\ell}=-\frac{\lambda_{20}}{2} \\
  E_{1221}&=-\sum_{\ell=1}^n (Q^{\ell}_{1211}\delta^{\ell}_{12}+Q^{\ell}_{1221}\delta^{\ell}_{22})h^{\ell}=-\sum_{\ell=1}^n (Q^{\ell}_{16}\delta^{\ell}_{xy}+Q^{\ell}_{66}\gamma^{\ell}_{y})h^{\ell}=-\frac{\lambda_{34}}{2} 
\end{align}
\end{subequations}
\paragraph{Matrix $\mymat{F}$}
Formulas~\eqref{eq:generalized stiffnesses1} and~\eqref{eq:deltabetaWoodcock} give:
\begin{gather}
  F_{\alpha\beta\mu\delta}=\int^{\zeta^n}_{\zeta^0} Q_{\alpha\beta\gamma\delta}\varphi_{\gamma\mu}(z)z\text{d}z=\sum_{\ell=1}^n Q^{\ell}_{\alpha\beta\gamma\delta}\left(\delta^{\ell}_{\gamma\mu} \beta^{\ell} h^{\ell} + \alpha_{\gamma\mu}^{\ell} \frac{(h^{\ell})^3}{12}\right)
\end{gather}
This formula gives Woodcock's coefficients $\lambda_i$ with $i\in (4,10,14,16,27,28,40,42,48,55,57,63)$. Let us finish the demonstration for some of them, changing the notations on the fly to fit Woodcock's coefficients:
\begin{subequations}
\begin{align}
  \nonumber
  F_{2222}&=\sum_{\ell=1}^n \left[Q^{\ell}_{2212}\left(\delta^{\ell}_{12} \beta^{\ell} h^{\ell} + \alpha_{12}^{\ell} \frac{(h^{\ell})^3}{12}\right)
	                               +Q^{\ell}_{2222}\left(\delta^{\ell}_{22} \beta^{\ell} h^{\ell} + \alpha_{22}^{\ell} \frac{(h^{\ell})^3}{12}\right)\right] \\
					&=\sum_{\ell=1}^n \left[Q^{\ell}_{26}\left(\delta^{\ell}_{xy} \beta^{\ell} h^{\ell} + \alpha_{xy}^{\ell} \frac{(h^{\ell})^3}{12}\right)
					                       +Q^{\ell}_{22}\left(\gamma^{\ell}_{y} \beta^{\ell} h^{\ell} + \alpha_{y}^{\ell} \frac{(h^{\ell})^3}{12}\right)\right] 
																=\frac{\lambda_{10}}{2} \\
  \nonumber
	F_{1122}&=\sum_{\ell=1}^n \left[Q^{\ell}_{1112}\left(\delta^{\ell}_{12} \beta^{\ell} h^{\ell} + \alpha_{12}^{\ell} \frac{(h^{\ell})^3}{12}\right)
	                               +Q^{\ell}_{1122}\left(\delta^{\ell}_{22} \beta^{\ell} h^{\ell} + \alpha_{22}^{\ell} \frac{(h^{\ell})^3}{12}\right)\right] \\
					&=\sum_{\ell=1}^n \left[Q^{\ell}_{16}\left(\delta^{\ell}_{xy} \beta^{\ell} h^{\ell} + \alpha_{xy}^{\ell} \frac{(h^{\ell})^3}{12}\right)
					                       +Q^{\ell}_{12}\left(\gamma^{\ell}_{y} \beta^{\ell} h^{\ell} + \alpha_{y}^{\ell} \frac{(h^{\ell})^3}{12}\right)\right] 
																=\frac{\lambda_{14}}{2} \\
  \nonumber
	F_{1121}&=\sum_{\ell=1}^n \left[Q^{\ell}_{1111}\left(\delta^{\ell}_{12} \beta^{\ell} h^{\ell} + \alpha_{12}^{\ell} \frac{(h^{\ell})^3}{12}\right)
	                               +Q^{\ell}_{2221}\left(\delta^{\ell}_{22} \beta^{\ell} h^{\ell} + \alpha_{22}^{\ell} \frac{(h^{\ell})^3}{12}\right)\right] \\
					&=\sum_{\ell=1}^n \left[Q^{\ell}_{11}\left(\delta^{\ell}_{xy} \beta^{\ell} h^{\ell} + \alpha_{xy}^{\ell} \frac{(h^{\ell})^3}{12}\right)
					                       +Q^{\ell}_{16}\left(\gamma^{\ell}_{y} \beta^{\ell} h^{\ell} + \alpha_{y}^{\ell} \frac{(h^{\ell})^3}{12}\right)\right] 
																=\frac{\lambda_{48}}{2} \\
  \nonumber
	F_{2212}&=\sum_{\ell=1}^n \left[Q^{\ell}_{2212}\left(\delta^{\ell}_{11} \beta^{\ell} h^{\ell} + \alpha_{11}^{\ell} \frac{(h^{\ell})^3}{12}\right)
	                               +Q^{\ell}_{2222}\left(\delta^{\ell}_{21} \beta^{\ell} h^{\ell} + \alpha_{21}^{\ell} \frac{(h^{\ell})^3}{12}\right)\right] \\
					&=\sum_{\ell=1}^n \left[Q^{\ell}_{26}\left(\gamma^{\ell}_{x} \beta^{\ell} h^{\ell} + \alpha_{x}^{\ell} \frac{(h^{\ell})^3}{12}\right)
					                       +Q^{\ell}_{22}\left(\delta^{\ell}_{yx} \beta^{\ell} h^{\ell} + \alpha_{yx}^{\ell} \frac{(h^{\ell})^3}{12}\right)\right] 
																=\frac{\lambda_{55}}{2} 
\end{align}
\end{subequations}
\paragraph{Matrix $\mymat{G}$}
Formulas~\eqref{eq:generalized stiffnesses1} and~\eqref{eq:delta2ampha2Woodcock} give:
\begin{align}
  G_{\nu\beta\mu\delta}&=\int^{\zeta^n}_{\zeta^0} Q_{\alpha\beta\gamma\delta}\varphi_{\alpha\nu}(z)\varphi_{\gamma\mu}(z)\text{d}z = \sum_{\ell=1}^n Q_{\alpha\beta\gamma\delta} 
	  \left(\delta_{\alpha\nu}\delta_{\gamma\mu}h^\ell + \alpha_{\alpha\nu}\alpha_{\gamma\mu}\frac{(h^\ell)^3}{12}\right)
\end{align}
This formula gives Woodcock's coefficients $\lambda_i$ with $i\in (2,8,17,23,24,31,43,50,58,65)$. Let us finish the demonstration for some of them, changing the notations on the fly to fit Woodcock's coefficients:
\begin{subequations}
\begin{align}
  \nonumber
  G_{1111}=\sum_{\ell=1}^n \bigg[&Q^{\ell}_{1111}\left(\delta^{\ell}_{11}\delta^{\ell}_{11} h^{\ell} + \alpha^{\ell}_{11}\alpha^{\ell}_{11} \frac{(h^{\ell})^3}{12}\right)
	                               +Q^{\ell}_{1121}\left(\delta^{\ell}_{11}\delta^{\ell}_{21} h^{\ell} + \alpha^{\ell}_{11}\alpha^{\ell}_{21} \frac{(h^{\ell})^3}{12}\right) \\
  \nonumber
	                              +&Q^{\ell}_{2111}\left(\delta^{\ell}_{21}\delta^{\ell}_{11} h^{\ell} + \alpha^{\ell}_{21}\alpha^{\ell}_{11} \frac{(h^{\ell})^3}{12}\right) 
	                               +Q^{\ell}_{2121}\left(\delta^{\ell}_{21}\delta^{\ell}_{21} h^{\ell} + \alpha^{\ell}_{21}\alpha^{\ell}_{21} \frac{(h^{\ell})^3}{12}\right)\bigg] \\
  \nonumber
					=\sum_{\ell=1}^n \bigg[&Q^{\ell}_{11}\left((\gamma^{\ell}_{x})^2 h^{\ell} + (\alpha^{\ell}_{x})^2 \frac{(h^{\ell})^3}{12}\right)
					                      +2Q^{\ell}_{16}\left(\gamma^{\ell}_{x} \delta^{\ell}_{yx} h^{\ell} + \alpha^{\ell}_{x}\alpha^{\ell}_{yx} \frac{(h^{\ell})^3}{12}\right) \\
														    +&Q^{\ell}_{66}\left((\delta^{\ell}_{yx})^2 h^{\ell} + (\alpha^{\ell}_{yx})^2 \frac{(h^{\ell})^3}{12}\right)\bigg] =\lambda_{2} \\
  \nonumber
  G_{1212}=\sum_{\ell=1}^n \bigg[&Q^{\ell}_{1212}\left(\delta^{\ell}_{11}\delta^{\ell}_{11} h^{\ell} + \alpha^{\ell}_{11}\alpha^{\ell}_{11} \frac{(h^{\ell})^3}{12}\right)
	                               +Q^{\ell}_{1222}\left(\delta^{\ell}_{11}\delta^{\ell}_{21} h^{\ell} + \alpha^{\ell}_{11}\alpha^{\ell}_{21} \frac{(h^{\ell})^3}{12}\right) \\
  \nonumber
	                              +&Q^{\ell}_{2212}\left(\delta^{\ell}_{21}\delta^{\ell}_{11} h^{\ell} + \alpha^{\ell}_{21}\alpha^{\ell}_{11} \frac{(h^{\ell})^3}{12}\right) 
	                               +Q^{\ell}_{2222}\left(\delta^{\ell}_{21}\delta^{\ell}_{21} h^{\ell} + \alpha^{\ell}_{21}\alpha^{\ell}_{21} \frac{(h^{\ell})^3}{12}\right)\bigg] \\
  \nonumber
					=\sum_{\ell=1}^n \bigg[&Q^{\ell}_{66}\left((\gamma^{\ell}_{x})^2 h^{\ell} + (\alpha^{\ell}_{x})^2 \frac{(h^{\ell})^3}{12}\right)
					                      +2Q^{\ell}_{26}\left(\gamma^{\ell}_{x} \delta^{\ell}_{yx} h^{\ell} + \alpha^{\ell}_{x}\alpha^{\ell}_{yx} \frac{(h^{\ell})^3}{12}\right) \\
														    +&Q^{\ell}_{22}\left((\delta^{\ell}_{yx})^2 h^{\ell} + (\alpha^{\ell}_{yx})^2 \frac{(h^{\ell})^3}{12}\right)\bigg] 
														    =\lambda_{23} \\
  \nonumber
  G_{1221}=\sum_{\ell=1}^n \bigg[&Q^{\ell}_{1211}\left(\delta^{\ell}_{11}\delta^{\ell}_{12} h^{\ell} + \alpha^{\ell}_{11}\alpha^{\ell}_{12} \frac{(h^{\ell})^3}{12}\right)
	                               +Q^{\ell}_{1221}\left(\delta^{\ell}_{11}\delta^{\ell}_{22} h^{\ell} + \alpha^{\ell}_{11}\alpha^{\ell}_{22} \frac{(h^{\ell})^3}{12}\right) \\
  \nonumber
	                              +&Q^{\ell}_{2211}\left(\delta^{\ell}_{21}\delta^{\ell}_{12} h^{\ell} + \alpha^{\ell}_{21}\alpha^{\ell}_{12} \frac{(h^{\ell})^3}{12}\right) 
	                               +Q^{\ell}_{2221}\left(\delta^{\ell}_{21}\delta^{\ell}_{22} h^{\ell} + \alpha^{\ell}_{21}\alpha^{\ell}_{22} \frac{(h^{\ell})^3}{12}\right)\bigg] \\
  \nonumber
					=\sum_{\ell=1}^n \bigg[&Q^{\ell}_{16}\left(\gamma^{\ell}_{x}  \delta^{\ell}_{xy} h^{\ell} + \alpha^{\ell}_{x} \alpha^{\ell}_{xy} \frac{(h^{\ell})^3}{12}\right)
					                       +Q^{\ell}_{66}\left(\gamma^{\ell}_{x}  \gamma^{\ell}_{y}  h^{\ell} + \alpha^{\ell}_{x} \alpha^{\ell}_{y}  \frac{(h^{\ell})^3}{12}\right) \\
					                      +&Q^{\ell}_{12}\left(\delta^{\ell}_{yx} \delta^{\ell}_{xy} h^{\ell} + \alpha^{\ell}_{yx}\alpha^{\ell}_{xy} \frac{(h^{\ell})^3}{12}\right)
														     +Q^{\ell}_{26}\left(\delta^{\ell}_{yx} \gamma^{\ell}_{y}  h^{\ell} + \alpha^{\ell}_{yx}\alpha^{\ell}_{y}  \frac{(h^{\ell})^3}{12}\right)\bigg] 
														    =\frac{\lambda_{31}}{2} \\
  \nonumber
  G_{1121}=\sum_{\ell=1}^n \bigg[&Q^{\ell}_{1111}\left(\delta^{\ell}_{11}\delta^{\ell}_{12} h^{\ell} + \alpha^{\ell}_{11}\alpha^{\ell}_{12} \frac{(h^{\ell})^3}{12}\right)
	                               +Q^{\ell}_{1121}\left(\delta^{\ell}_{11}\delta^{\ell}_{22} h^{\ell} + \alpha^{\ell}_{11}\alpha^{\ell}_{22} \frac{(h^{\ell})^3}{12}\right) \\
  \nonumber
	                              +&Q^{\ell}_{2111}\left(\delta^{\ell}_{21}\delta^{\ell}_{12} h^{\ell} + \alpha^{\ell}_{21}\alpha^{\ell}_{12} \frac{(h^{\ell})^3}{12}\right) 
	                               +Q^{\ell}_{2121}\left(\delta^{\ell}_{21}\delta^{\ell}_{22} h^{\ell} + \alpha^{\ell}_{21}\alpha^{\ell}_{22} \frac{(h^{\ell})^3}{12}\right)\bigg] \\
  \nonumber
					=\sum_{\ell=1}^n \bigg[&Q^{\ell}_{11}\left(\gamma^{\ell}_{x}  \delta^{\ell}_{xy} h^{\ell} + \alpha^{\ell}_{x} \alpha^{\ell}_{xy} \frac{(h^{\ell})^3}{12}\right)
					                       +Q^{\ell}_{16}\bigg(\gamma^{\ell}_{x}  \gamma^{\ell}_{y}  h^{\ell} + \alpha^{\ell}_{x} \alpha^{\ell}_{y}  \frac{(h^{\ell})^3}{12} \\
					                      +&\delta^{\ell}_{yx} \delta^{\ell}_{xy} h^{\ell} + \alpha^{\ell}_{yx}\alpha^{\ell}_{xy} \frac{(h^{\ell})^3}{12}\bigg)
														     +Q^{\ell}_{66}\left(\delta^{\ell}_{yx} \gamma^{\ell}_{y}  h^{\ell} + \alpha^{\ell}_{yx}\alpha^{\ell}_{y}  \frac{(h^{\ell})^3}{12}\right)\bigg] 
																=\frac{\lambda_{50}}{2} 
\end{align}
\end{subequations}
Note the specific structure of these coefficients, some of them involving $3$ different stiffness tensor components (with also different coefficients structures) like~$\lambda_{23}$ and~$\lambda_{50}$, and some of them involving $4$ different stiffness tensor components, like~$\lambda_{31}$. 
\paragraph{Matrix $\mymat{H}$} Equations~\eqref{eq:generalized stiffnesses1}, \eqref{eq:def_deriv_warp} and~\eqref{eq:blocA} give:
\begin{align}
  H_{\alpha3\beta3}&=\int^{\zeta^n}_{\zeta^0} \varphi'_{\gamma\alpha}(z) C_{\gamma3\delta3} \varphi'_{\delta\beta}(z) \text{d}z = \sum_{\ell=1}^n C^\ell_{\gamma3\delta3} 
	  \alpha^\ell_{\gamma\alpha}\alpha^\ell_{\delta\beta}h^\ell
\end{align}
This formula gives Woodcock's coefficients $\lambda_i$ with $i\in (37,38,69)$. Let us finish the demonstration for the $\lambda_{69}$:
\begin{align}
  \nonumber
  H_{1323}=\sum_{\ell=1}^n \bigg[&C^{\ell}_{1313}\left(\alpha^{\ell}_{11}\alpha^{\ell}_{12} h^{\ell}\right)
	                               +C^{\ell}_{2313}\left(\alpha^{\ell}_{21}\alpha^{\ell}_{12} h^{\ell}\right)
	                               +C^{\ell}_{1323}\left(\alpha^{\ell}_{11}\alpha^{\ell}_{22} h^{\ell}\right) 
	                               +C^{\ell}_{2323}\left(\alpha^{\ell}_{21}\alpha^{\ell}_{22} h^{\ell}\right)\bigg] \\
					=\sum_{\ell=1}^n \bigg[&C^{\ell}_{55} \alpha^{\ell}_{x}\alpha^{\ell}_{xy} h^{\ell}
					                       +C^{\ell}_{45} (\alpha^{\ell}_{yx}\alpha^{\ell}_{xy} + \alpha^{\ell}_{x}\alpha^{\ell}_{y}) h^{\ell}
														     +C^{\ell}_{44} \alpha^{\ell}_{yx} \alpha^{\ell}_{y} h^{\ell}\bigg] = \frac{\lambda_{69}}{2}
\end{align}
\paragraph{Matrix $\mymat{\boldsymbol{\Xi}}$} Equations~\eqref{eq:generalized_mass},~\eqref{eq:intdz},~\eqref{eq:intzdz},~\eqref{eq:intz2dz},~\eqref{eq:deltaWoodcock},~\eqref{eq:deltabetaWoodcock} and~\eqref{eq:delta2ampha2Woodcock} permit to write:
\begin{align}
  R&=\int^{\zeta^n}_{\zeta^0}\rho(z)\text{d}z=\sum_{\ell=1}^n \rho^{\ell} h^{\ell}=\delta_{3}=\delta_{9}=\delta_{13} \\
  S&=\int^{\zeta^n}_{\zeta^0}\rho(z)z\text{d}z=-\sum_{\ell=1}^n \rho^{\ell} \beta^{\ell} h^{\ell}=-\delta_{5}=-\delta_{11} \\
  T&=\int^{\zeta^n}_{\zeta^0}\rho(z)z^2\text{d}z=\sum_{\ell=1}^n \rho^{\ell} \left(\frac{1}{12}(h^{\ell})^3+h^{\ell}(\beta^{\ell})^2\right)=\delta_{1}=\delta_{7} \\
  U_{\alpha\beta}&=\int^{\zeta^n}_{\zeta^0}\rho(z)\varphi_{\alpha\beta}\text{d}z=-\sum_{\ell=1}^n \rho^{\ell} \delta^{\ell}_{\alpha\beta} h^{\ell} \quad \text{ are the }-\frac{\delta_i}{2} \text{ for } i \in \{6,12,17,18\} \\
  V_{\alpha\beta}&=\int^{\zeta^n}_{\zeta^0}\rho(z)\varphi_{\alpha\beta}z\text{d}z=\sum_{\ell=1}^n \rho^{\ell} \left(\delta^{\ell}_{\alpha\beta} \beta^{\ell} h^{\ell} + \alpha_{\alpha\beta}^{\ell} \frac{(h^{\ell})^3}{12}\right) \quad \text{ are the } \frac{\delta_i}{2} \text{ for } i \in \{4,10,15,16\} \\
  W_{\alpha\beta}&=\int^{\zeta^n}_{\zeta^0}\rho(z)\varphi_{\mu\alpha}\varphi_{\mu\beta}\text{d}z=\sum_{\ell=1}^n \rho^{\ell} \left(\delta_{\mu\alpha}\delta_{\mu\beta}h^\ell + \alpha_{\mu\alpha}\alpha_{\mu\beta}\frac{(h^\ell)^3}{12}\right) \quad \text{ are the } \delta_i \text{ for } i \in \{2,8\} \text{ and } \frac{\lambda_{14}}{2}
\end{align}